%% file: score_paper.tex
\documentclass{article}

\input{Setup/preamble.tex}
\input{Setup/math.tex}

\input{General/front_page.tex} 

\begin{document}
\maketitle

\begin{abstract}
	\input{General/abstract.tex}
\end{abstract}

\input{General/keywords.tex}

\section{Introduction}
\input{General/Introduction.tex}

\section{Background and Related Work}
\subsection{Riemannian Geometry and Notation}
\input{Sections/Background/riemannian_geometry}
\subsection{Means on Riemannian Manifolds} \label{sec:means}
\input{Sections/Background/means}

\section{Estimating of Diffusion t-Means using Score Matching}
\input{Sections/estimation/score_matching}
\section{Experiments and Applications}
\input{Sections/experiments}
\section{Conclusion}
\input{General/conclusion}

\clearpage
\small
\bibliographystyle{plainnat} 
\bibliography{score_paper}  

\clearpage
\section*{Appendix}
\begin{appendix}
    \section{Sampling Riemannian Brownian Motion} \label{ap:sampling}
    \input{General/appendix/sampling}
    \section{Riemannian Manifolds} \label{ap:manifolds}
    \input{General/appendix/manifolds}
    \section{Heat Kernel} \label{ap:heat_kernel}
    \input{General/appendix/heat_kernels}
    \section{Data} \label{ap:data}
    \input{General/appendix/data}
    \section{Implementation} \label{ap:implementation}
    \input{General/appendix/implementation}
\end{appendix}







\end{document}

%% file: Setup/preamble.tex
\usepackage{arxiv}

\usepackage[utf8]{inputenc} 
\usepackage[T1]{fontenc}    
\usepackage{hyperref}       
\usepackage{url}            
\usepackage{booktabs}       
\usepackage{amsfonts}       
\usepackage{nicefrac}       
\usepackage{microtype}      
\usepackage{lipsum}		
\usepackage{graphicx}
\usepackage{natbib}
\usepackage{doi}
\usepackage{amsmath}
\usepackage{subcaption}
\usepackage{graphicx}
\usepackage{wrapfig}
\usepackage{ragged2e}
\usepackage[ruled,vlined]{algorithm2e}
\usepackage{graphicx,lipsum,wrapfig}

%% file: Setup/math.tex
\usepackage[english]{babel}
\usepackage{amsthm}
\usepackage{mathtools}
\usepackage[ruled,vlined]{algorithm2e}
\SetKwRepeat{Do}{do}{while}

\theoremstyle{plain}

\theoremstyle{definition}

\newcommand{\norm}[1]{\left\lVert#1\right\rVert}
\DeclareMathOperator*{\argmin}{arg\,min}
\DeclareMathOperator*{\argmax}{arg\,max}
\DeclarePairedDelimiterX{\inp}[2]{\langle}{\rangle}{#1, #2}

\newcommand\restr[2]{{
  \left.\kern-\nulldelimiterspace 
  #1 
  \vphantom{\big|} 
  \right|_{#2} 
  }}
  
\newcommand{\dist}[0]{\mathrm{dist}} 
\newcommand{\dif}[0]{\mathrm{d}} 

%% file: General/front_page.tex
\title{Score Matching Riemannian Diffusion Means}

\author{Frederik Möbius Rygaard \\
	Technical University of Denmark\\
	\texttt{fmry@dtu.dk} \\
	\And
	Steen Markvorsen \\
	Technical University of Denmark\\
	\texttt{stema@dtu.dk} \\ \\
        \And
	Søren Hauberg \\
	Technical University of Denmark\\
	\texttt{sohau@dtu.dk} \\
        \And
	Stefan Sommer \\
	University of Copenhagen\\
	\texttt{sommer@di.ku.dk} \\
}

\date{}

\hypersetup{
pdftitle={Score Matching Riemannian Diffusion Means},
pdfsubject={Estimation of Diffusion t-Means using Score Matching},
pdfauthor={Frederik Möbius Rygaard, Steen Markvorsen, Søren Hauberg, Stefan Sommer},
pdfkeywords={Riemannian Manifolds, Diffusion t-means, Intrinsic Statistics, Score Matching},
}

%% file: General/abstract.tex
Estimating means on Riemannian manifolds is generally computationally expensive because the Riemannian distance function is not known in closed-form for most manifolds. To overcome this, we show that Riemannian diffusion means can be efficiently estimated using score matching with the gradient of Brownian motion transition densities using the same principle as in Riemannian diffusion models. Empirically, we show that this is more efficient than Monte Carlo simulation while retaining accuracy and is also applicable to learned manifolds. Our method, furthermore, extends to computing the Fréchet mean and the logarithmic map for general Riemannian manifolds. We illustrate the applicability of the estimation of diffusion mean by efficiently extending Euclidean algorithms to general Riemannian manifolds with a Riemannian $k$-means algorithm and maximum likelihood Riemannian regression.\looseness=-1

%% file: General/keywords.tex
\keywords{Riemannian Manifolds \and Diffusion t-means \and Intrinsic Statistics \and Score Matching}

%% file: General/Introduction.tex
\begin{wrapfigure}[11]{r}{0.5\textwidth}
    \vspace{-3.5em}
    \centering
    \includegraphics[width=0.45\textwidth]{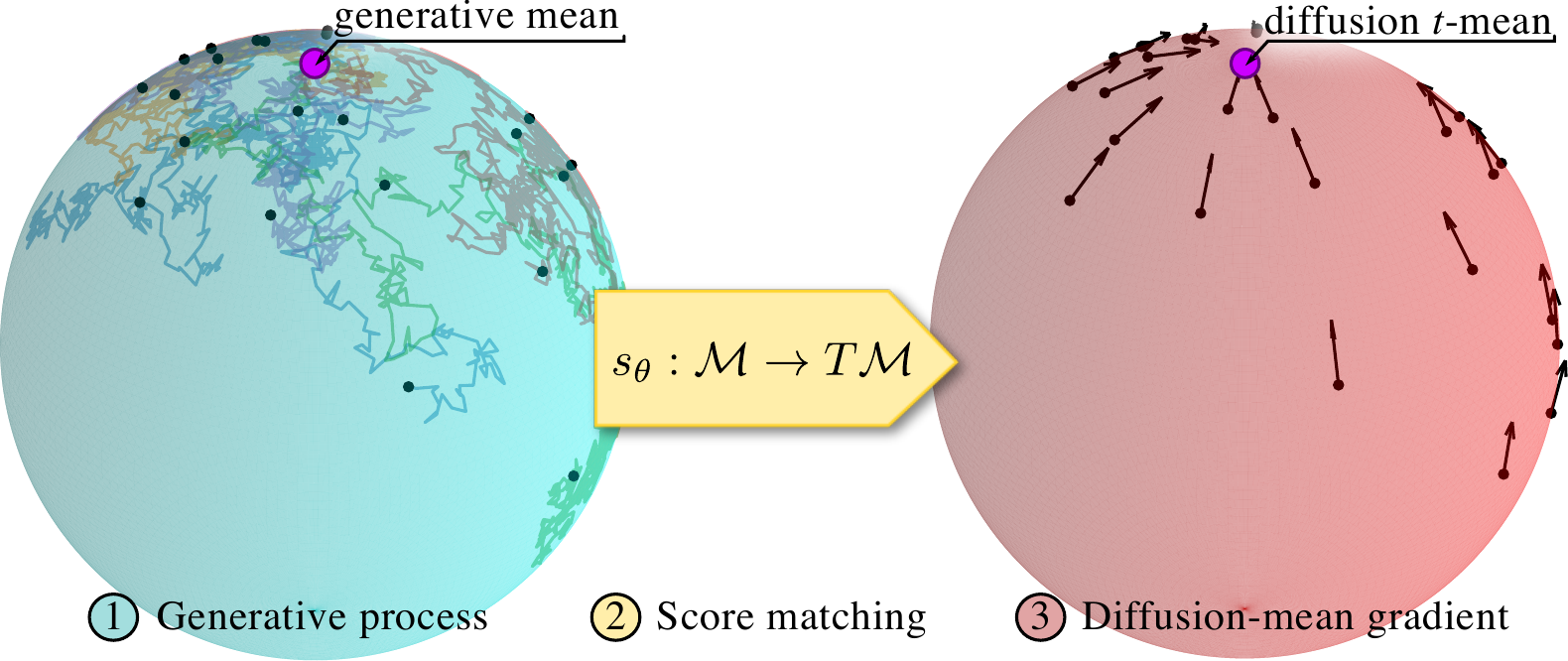}
    \caption{The diffusion mean is estimated by learning the gradients of the log-transition density. (1) Sample points from a Riemannian Brownian motion are (2) passed through a neural network minimizing the score matching loss and (3) outputs the gradients used to estimate the diffusion $t$-mean.}
    \label{fig:score_illustration}
    \vspace{3.0em}
\end{wrapfigure}
The \emph{average} of a dataset is arguably one of the most essential statistical quantities of interest, and it forms a building block for more elaborate models. For data residing on a Riemannian manifold, the most common definition of an average is the \emph{Fr\'echet mean}, which minimizes the sum of squared geodesic distances to data \citep{frechet1948},
\begin{align}
  \mu^{\text{Fr\'echet}}(x_{1:N}) &:= \argmin_{\mu \in \mathcal{M}} \sum_{n=1}^N \dist^2(x_n, \mu),
  \label{eq:Frmean_sample}
\end{align}
where $\dist(\cdot, \cdot)$ is the distance along the manifold, $\mathcal{M}$. The Fr\'echet mean is not the only generalization of the Euclidean mean value to non-Euclidean domains, and it has properties, which makes it worthwhile to consider alternatives: \emph{Computationally}, the Fr\'echet mean is expensive when the manifold distance function is not known in closed-form, which is the case for all but only a selected set of manifolds. \emph{Theoretically}, the Fr\'echet mean is only a maximum likelihood estimate when the underlying manifold is restricted to being \emph{symmetric} \citep{fletcher:geodesicregression:2013}, which is a very restrictive assumption satisfied by the simplest of manifolds, and the sample Fr\'echet mean can exhibit slow convergence to the Fr\'echet mean of the data distribution \citep{eltzner2022diffusion}. These properties have recently motivated the construction of the \emph{diffusion $t$-mean} \citep{eltzner2022diffusion}. The \emph{diffusion t-mean} is the maximum likelihood estimator for the transition density of a Brownian motion on the manifold \citep{eltzner2022diffusion},
\begin{align}
  \mu_{t}^{\text{Diffusion}}(x_{1:N}) &:= \argmax_{\mu \in \mathcal{M}} \sum_{n=1}^N \log p_{t}(x_n, \mu),
  \label{eq:diffmean}
\end{align}
where $p$ is the transition density with diffusion time $t$. The diffusion mean is per definition a maximum likelihood estimate, and, in many situations, the sample diffusion mean avoids the slow convergence to the distribution mean. However, the construction comes with new computational difficulties. For general Riemannian manifolds, the transition density of a Brownian motion corresponds to the minimal heat kernel for which analytic expressions are rarely available. This implies that evaluating the log-likelihood objective function is intractable. \citet{sommer2017bridge} suggest using \emph{bridge sampling} to approximate the log-likelihood, but such approximations are computationally expensive.
\begin{figure}[t!]
    \centering
    \includegraphics[width=1.0\textwidth]{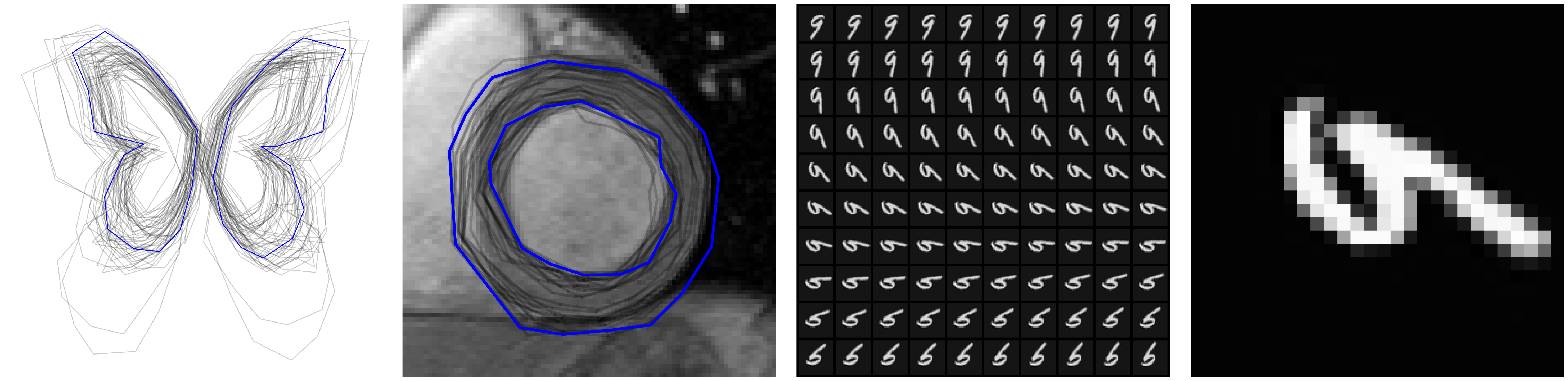}
    \caption{Estimated values of the diffusion mean using score matching for two different manifolds. The two left most plots show the diffusion mean (blue) for a shape space with $30 \times 2$ landmarks on butterfly and cardiac data (black), respectively. The two right most plots shows a learned manifold using a Gaussian process $\mathcal{GP}$ with corresponding expected metric for a rotated MNIST image. The first plot shows the reconstructed rotated MNIST data for half a rotation, while the right most plot shows the estimated diffusion mean for half a rotation.}
    \vspace{-1.0em}
    \label{fig:score_real_data}
\end{figure}

\textbf{In this paper}, we show that \emph{Riemannian score matching} \citep{debortoli2022riemannian} can be used to efficiently estimate the diffusion t-mean for general manifolds that scale linearly with the manifold dimension. This is a vast improvement over previous approaches with cubic complexity. We further show that the Fr\'echet mean \eqref{eq:Frmean_sample} can be estimated using the limit of the scores. Empirically, we verify that our method is more efficient than Monte Carlo bridge sampling, which is the only available alternative for general manifolds and apply it to learned manifolds. We further illustrate how the efficient estimation of the diffusion mean allows us to perform Riemannian $k$-means clustering and maximum likelihood Riemannian regression. In general, the presented algorithms provide a practical approach to compute statistics on general Riemannian manifolds efficiently.

%% file: Sections/Background/riemannian_geometry.tex
\textbf{A Riemannian manifold}, $(\mathcal{M},g)$, is a differentiable manifold equipped with a Riemannian metric, $g$, which defines a smoothly varying inner product, $\langle \cdot, \cdot \rangle_{x}$ on each tangent space, $T_{x}\mathcal{M}$ for $x \in \mathcal{M}$. The tangent space, $T_{x}\mathcal{M}$, is a vector space consisting of all tangent vectors for some curve $\gamma: (-\epsilon, \epsilon) \rightarrow \mathcal{M}$ with $\gamma(0) = x \in \mathcal{M}$ \citep{do1992riemannian}. The tangent bundle, $T\mathcal{M}$ is the disjoint union of the tangent spaces, i.e.\@ $T\mathcal{M} = \bigsqcup_{x \mathcal{M}} T_{x}\mathcal{M}$. Some simple examples of Riemannian manifolds are the $m$-sphere, $\mathbb{S}^{m} = \{x \in \mathbb{R}^{m+1}\,|\,\|x\| = 1\}$ with metric inherited from the ambient Euclidean space; the torus $\mathbb{T}^{m}$, which is the direct product of $m$ copies of $\mathbb{S}^{1}$ equipped with e.g.\@ a flat metric, and also, of course, Euclidean space itself, $\mathbb{R}^{m}$.\looseness=-1

\textbf{A geodesic} is a curve, $\gamma: U \subseteq \mathbb{R}_{+} \rightarrow \mathcal{M}$ that locally minimizes the curve length, $\mathcal{L}(\gamma) = \int_{0}^{1} \langle\dot{\gamma}(t), \dot{\gamma}(t)\rangle_{\gamma(t)} \,\dif t$. The distance on a Riemannian manifold is defined as the length of the geodesic connecting two points, i.e. $\dist(x,y) = \min_{\gamma} \mathcal{L}(\gamma)$ for $x,y \in \mathcal{M}$. We will assume that the manifolds is geodesically complete in the sense that between any pair of points, $x, y \in \mathcal{M}$, there exists at least one length-minimizing geodesic, $\gamma$, with $\gamma(0) = x$ and $\gamma(1) = y$. In practice, a geodesic can be found by either minimizing the energy functional, $\mathcal{E}(\gamma) = \frac{1}{2}\int_{0}^{1} \langle \dot{\gamma}(t), \dot{\gamma}(t)\rangle_{\gamma(t)} \,\dif t$ or by solving the following, usually nonlinear, \textsc{ode}
\begin{equation}
    \frac{\dif x^{k}}{\dif t^{2}}+\Gamma_{ij}^{k}\frac{\dif x^{i}}{\dif t}\frac{\dif x^{j}}{\dif t} = 0,
    \label{eq:geodesic_ode}
\end{equation} 
written in a local coordinate chart in Einstein notation, where $\Gamma_{ij}^{k}$ denotes the Christoffel symbols. The exponential map, $\text{Exp}_{x}: \mathcal{T}_{x}\mathcal{M} \rightarrow \mathcal{M}$, is given by $\text{Exp}_{x}(v) = \gamma(1)$, where $\gamma$ is a geodesic with $\gamma(0)=x$ and $\dot{\gamma} = v$. It can be shown that the exponential map is a diffeomorphism in a star-shaped neighborhood, $\mathcal{D}(x) \subseteq T_{x}\mathcal{M}$ \citep{do1992riemannian}. Since $\mathcal{M}$ is assumed geodesically complete, the curve $\gamma_{x}(t) = \text{Exp}_{x}(t v)$ is either a length-minimizing up to a point $t_{0}$ or is length minimizing for all $t \in [0, \infty[$ \citep{pennec2006statriemann}. In the first case $t_{0}$ is called a \emph{cut point}, and the set of all cut points, $\mathcal{C}(x)$, is called the \emph{cut locus}. Within the cut locus, the inverse of the exponential map is given by the logarithmic map, $\text{Log}_{x}: \mathcal{M} \setminus \mathcal{C}(x) \rightarrow T_{x}\mathcal{M}$. A measure on a Riemannian manifold, $(\mathcal{M},g)$, is given by infinitesimal volume on each tangent space $\dif \mathcal{M}(x) = \sqrt{|g(x)|}\,\dif x$, where $|g(x)|$ denotes the determinant of the local representation of the metric. The Riemannian divergence, $\mathrm{div}$, is defined in a local coordinates system as $\mathrm{div}(V) = \frac{\partial V^{m}}{\partial x^{m}}+V^{k}\frac{\partial \log \sqrt{\log |g|}}{\partial x^{k}}$ for a vector field, $V: \mathcal{M} \rightarrow T\mathcal{M}$.
\begin{figure}[t!]
    \centering
    \includegraphics[width=1.0\textwidth]{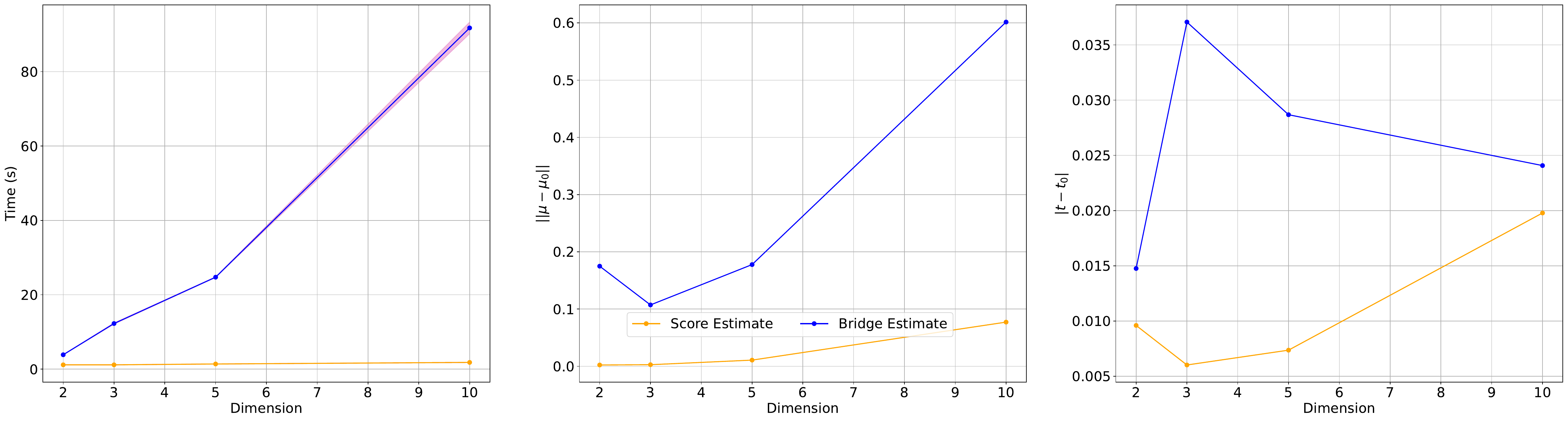}
    \caption{Run-Time and error comparison for $\mathbb{S}^{d}$ with $d=2,3,5,10$ between score based gradient descent (yellow) and bridge sampling (blue) to estimate the diffusion mean. The left most plot shows the run-time over dimension, while the center and right plot show the error of the diffusion mean and variance estimate, respectively. The estimates are based on $1,000$ samples of Brownian motion at the north pole for diffusion time $t=0.5$.}
    \vspace{-1.0em}
    \label{fig:timing_Sphere}
\end{figure}

\textbf{Riemannian Brownian motion} is a stochastic process on $\mathcal M$ that, when started at $x\in\mathcal M$ has the minimal heat kernel $p_{t}(x,\cdot)$ as transition density for $t \in \mathbb{R}_{+}$. The heat kernel is the solution to the \textsc{pde}
\begin{equation}
    \partial_{t}p_{t}(x,y) = \frac{1}{2}\Delta_{x}p_{t}(x,y),
    \label{eq:heat_kernel}
\end{equation}
where $\Delta$ denotes the Laplace-Beltrami operator. Note that there are other equivalent ways of defining Riemannian Brownian motion, see e.g.\@ \citet{hsustochastic} for details. We will assume that the manifold $\mathcal{M}$ is stochastic complete such that there exists a solution to eq.~\ref{eq:heat_kernel}, and that $\mathcal{M}$ has bounded sectional curvature to ensure the existence of a bounded minimal heat kernel $p$ \citep{eltzner2022diffusion}. Further, we will assume that $\mathcal{M}$ is orientable and boundaryless without any singularities or boundary points in accordance with \citet{debortoli2022riemannian} in order to sample paths of Brownian motions.

%% file: Sections/Background/means.tex
\textbf{The Fr\'echet mean} minimizes the expected squared distances with respect to the Riemannian metric \citep{frechet1948} generalizing the Euclidean mean, which minimizes the expected squared distances with respect to the Euclidean norm.
\begin{equation}
    \mu^{\text{Fr\'echet}}(X) := \argmin_{y \in \mathcal{M}} \mathbb{E}\left[\dist^{2}(X,y)\right],
    \label{eq:Frmean_expectation}
\end{equation}
for some random variable, $X$, on $\mathcal{M}$ with the probability space $(\Omega, \mathcal{F}, \mathbb{P})$. Unlike the Euclidean mean, the Fr\'echet mean is not necessarily unique. Depending on the curvature of $\mathcal{M}$, $\mu^{\text{Fr\'echet}}$ may consist of multiple points and thus being a \emph{mean set} \citep{Karcher1977RiemannianCO}. For a given set of observations $\{x_{i}\}_{i=1}^{n} \subset \mathcal{M}$, the sample estimator of the Fr\'echet mean is given by eq.~\ref{eq:Frmean_sample}. The Fr\'echet mean can be numerically estimated using a gradient descent approach with gradient $\nabla_{y} \mathbb{E}\left[\dist^{2}(x,y)\right] = \mathbb{E}\left[-2 \text{Log}_{y}(x)\right]$ \citep{pennec2006statriemann}. With this approach, computing the sample Fr\'echet mean requires estimating the logarithmic map, which is numerically expensive, since it involves minimizing either the energy functional or solving the \textsc{ode} \eqref{eq:geodesic_ode} as a boundary value problem. Other approaches to estimate the Fr\'echet mean without using logarithmic maps have been investigated, where \citet{lou2021differentiating} investigates using direct differentiation of the objective function focusing on Hyperbolic spaces. However, these methods are computationally expensive or rely on closed-form expressions on the manifold.
\begin{figure}[t!]
    \centering
    \includegraphics[width=1.0\textwidth]{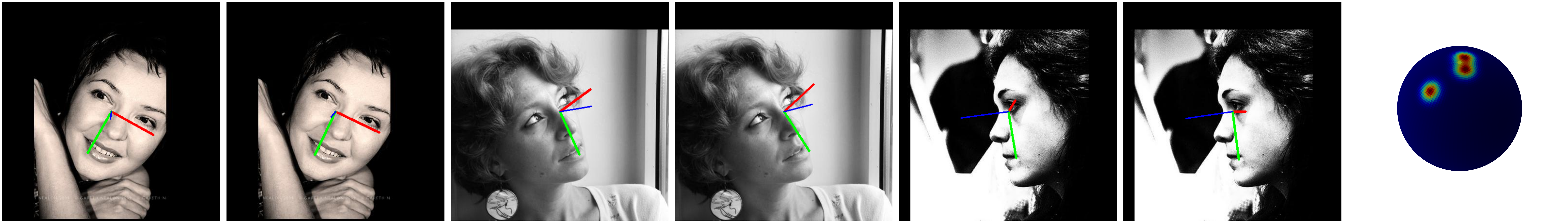}
    \caption{The estimated head-pose position for AFLW-2000 dataset, where the arrows correspond to the head pose position. The left image is the true labeled direction, while the right image is the prediction. Using only the roll and yaw of the Euler angles, we plot the mixture of the maximum likelihood neural regression on $\mathbb{S}^{2}$ using the estimated un-normalized heat kernel on $\mathbb{S}^{2}$.}
    \vspace{-1.5em}
    \label{fig:head_pose_s3}
\end{figure}
\textbf{The diffusion $t$-mean} is inspired by the Euclidean maximum likelihood estimation. Let $p$ denote the minimal heat kernel solving eq.~\ref{eq:heat_kernel} for a given Riemannian manifold, $(\mathcal{M},g)$, then the diffusion mean is defined as \citep{eltzner2022diffusion}
\begin{equation}
    \mu_{t}^{\text{Diffusion}}(X) = \argmax_{y \in \mathcal{M}} \mathbb{E}\left[ \log p_{t}(X,y)\right],
    \label{eq:diffusion_mean}
\end{equation}
Note that $\mu_{t}^{\text{Diffusion}}$ can in general be a set similar to the Fr\'echet mean and is denoted as the \emph{diffusion $t$-mean set} \citep{eltzner2022diffusion}. In Euclidean space, the distribution of the Brownian motion at time $t$ is a normal distribution with variance $t$. The diffusion mean can thus be interpreted as the center of a non-Euclidean normal distribution fitted to the data. The time $t$ can be fitted to the data as well by minimizing the negative log-likelihood with respect to $t$. This optimal $t$ is called the \emph{diffusion variance}. A sample estimate of the log-likelihood is given in eq.~\ref{eq:diffmean}. The diffusion mean can be estimated using Monte Carlo simulation \citep{sommer2017bridge}, which uses samples from Brownian Bridges computed as a guided process \citep{DELYON20061660, Papaspiliopoulos2012}. Further improvements on the Monte Carlo simulation have been proposed by \citet{buiInferencePartiallyObserved2022} and \citet{jensensommer2022}. However, estimating the likelihood using Monte Carlo simulation requires computing the Logarithmic map making it numerically expensive, when there is no closed-form expression available. Unlike this, our method does not rely on closed-form expressions and is applicable to any dataset once the scores have been trained for a fixed manifold independent of the observations.\looseness=-1

\begin{wrapfigure}{r}{0.5\textwidth}
    \vspace{-1.0em}
    \centering
    \includegraphics[width=0.45\textwidth]{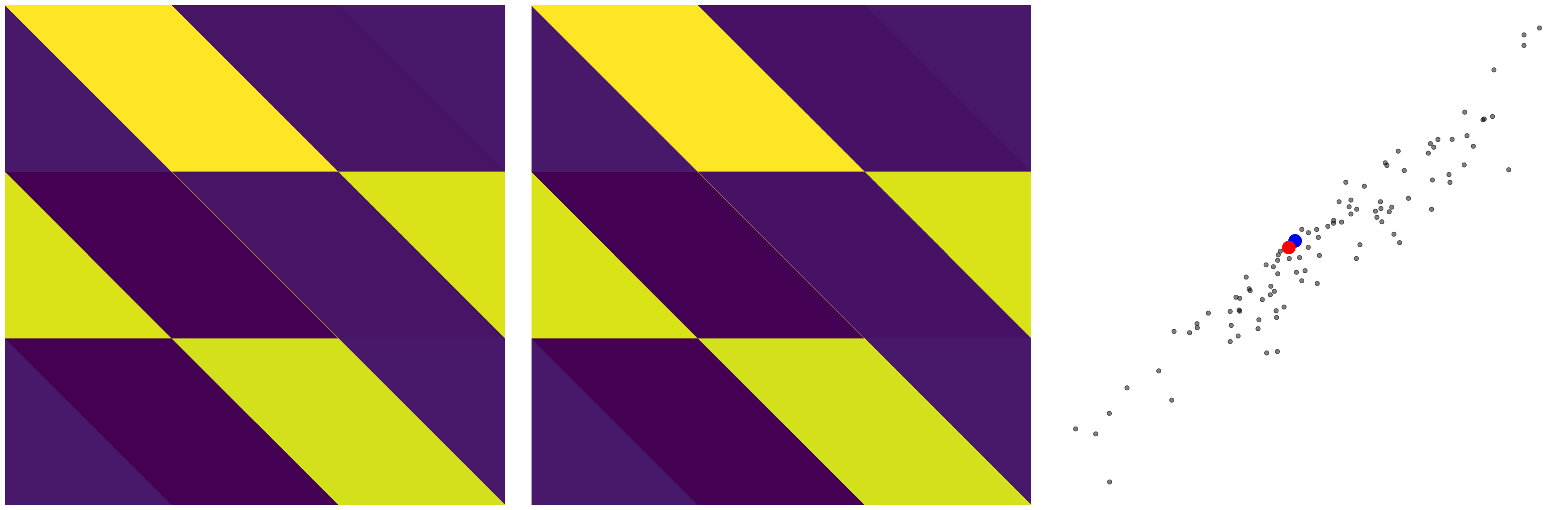}
    \caption{The estimated Fréchet mean for DTI tensor images corresponding to elements in $\mathcal{P}(3)$ using the time limit of the score. The two leftmost plots show the the Fr\^echet mean estimated using scores and with the Logarithmic map, respectively, while the right shows the two leading eigenvalues of the observations (black) as well as the approximation using scores (red) and ground truth (blue).}
    \label{fig:score_frechet_spdn}
    \vspace{-1.5em}
\end{wrapfigure}

\paragraph{Why use diffusion means?} Both the Fréchet mean and diffusion mean are intrinsic means, which take the underlying manifold structure into account rather than using extrinsic methods that can be computationally difficult to project back to the manifold and be less accurate. The diffusion mean and Fréchet mean can be seen as generalizations of the usual Euclidean mean in the sense that both means coincide with the mean in Euclidean space. In terms of the convergence of the estimators for $N$ sample points, $\mu_{N}^{\text{Fréchet}}$ and $\mu_{t,N}^{\text{Diffusion}}$, \citet{eltzner2022diffusion} shows that the estimator of the diffusion mean for a fixed $t>0$ is strongly consistent in the sense of \citet{Ziezold1977}, i.e. $\cap_{n=1}^{\infty} \overline{\cup_{k=n}^{\infty} \mu_{t,k}^{\text{Diffusion}}} \subset \mu_{t}^{\text{Diffusion}}$ for almost all $\omega$ and in the sense of \citet{bpc2003}, i.e. if $\mathcal{M} \neq \emptyset$ and if for almost all $\omega$ and for all $\epsilon>0$ there exists an $N_{\epsilon, \omega}$ such that $\cup_{k=N}^{\infty} \mu_{t,k}^{\text{Diffusion}}(\omega) \subset B(\mathcal{M, \epsilon})$, where $B(M, \epsilon)$ is the union of all balls with radius $\epsilon$ around all points on $\mathcal{M}$. The estimator for the Fréchet mean is similarly a strongly consistent estimator with respect to the above conditions and satisfies them for the same conditions as the diffusion mean. In the time limit $t \rightarrow 0$, $\mu_{t,N}^{\text{Diffusion}} \rightarrow \mu_{N}^{\text{Fréchet}}$ \citep{eltzner2022diffusion}. The Fréchet mean can be `smeary' in the sense that the sample Fr\'echet mean can exhibit slower convergence to the distribution mean than the standard $\sqrt{N}$ rate in the Euclidean central limit theorem. When the diffusion mean and the diffusion variance are estimated simultaneously, the diffusion has less tendency to exhibit smeariness (it can be smeary in some but not all directions) thus requiring fewer samples for a given precision of the sample estimate \citep{eltzner2022diffusion}. 

\paragraph{Computational Cost} The cost of estimating Fréchet means and diffusion means is numerically inexpensive, if the distances or heat kernel, respectively, have closed-form for the given Riemannian manifold. However, this is the case only known for a selected set of Riemannian manifolds. In general, estimating distances on Riemannian manifolds involves either solving the \textsc{ode} \eqref{eq:geodesic_ode} as a boundary value problem or minimizing the energy functional, both of which are computationally expensive. The heat kernel, and thus diffusion means, can be computed using Monte Carlo estimation of the heat kernel with diffusion bridges \citep{sommer2017bridge}. The computational complexity of this depends on the precision of the Monte Carlo estimate and is thus not directly comparable to the case for the Fr\'echet mean. In practice, the two approaches are often similarly expensive. The approach proposed in this paper substantially improves this situation making it computationally feasible to compute statistics on general Riemannian manifolds for both the diffusion and Fr\'echet mean.
\clearpage

%% file: Sections/estimation/score_matching.tex
\begin{wrapfigure}{R}{0.5\textwidth}
    \vspace{-1.7em}
    \begin{minipage}{0.5\textwidth}
        \begin{algorithm}[H]
            \SetAlgoLined
            \textbf{Input}: $N_{\text{iter}}$, $x_{1:N}$, $t^{\text{init}}$, $\mu^{\text{init}}$, $\alpha$ \\
            \textbf{Output}: Diffusion mean and diffusion variance \\
            Learn neural network $s_{1}$ minimizing eq.~\ref{eq:score_loss_dsm} \\
            Set $\mu = \mu^{\text{init}}, t = t^{\text{init}}$ \\
            \For{$i=0,\dots,N_{\text{iter}}$} {
                $\mu \leftarrow \text{Exp}_{\mu} \left(- \alpha \frac{1}{N}\sum_{i} s_{1}(x_{i},\mu,t;\theta)\right)$ \\
                $t \leftarrow -\alpha \frac{1}{N}\sum_{i} \frac{\dif}{\dif t} f(x_{i},\mu,t;\theta)$ with $f$ given by eq.~\ref{eq:gradt_log}
            }
            return $\mu, t$ \\
            \caption{Estimating Diffusion t-Mean}
            \label{al:est_diffusion_mean}
        \end{algorithm}
    \end{minipage}
    \vspace{-1.7em}
\end{wrapfigure}
\paragraph{Estimating the diffusion $t$-mean.} In practice the diffusion $t$-mean in eq. \ref{eq:diffusion_mean} can be estimated by gradient descent, where the gradient is given by $\nabla_{y} \mathbb{E}\left[-\log p_{t}(X, y)\right] = \mathbb{E}\left[- \nabla_{y}\log p_{t}(X, y)\right]$. Similarly, we find the diffusion variance by gradient descent using $\partial_{t} \mathbb{E}\left[-\log p_{t}(X, y)\right] = \mathbb{E}\left[- \partial_{t}\log p_{t}(X, y)\right]$. The optimization scheme is outlined in algorithm \ref{al:est_diffusion_mean}. However, for general Riemannian manifolds the gradient is not available in closed-form expression. We will instead approximate the gradient using the score of Riemannian Brownian motion.
%
\paragraph{Estimating the score.}Consider a Riemannian Brownian Motion, $W_{t}^{\mathcal{M}}$, with transition density $p_{t}(x,y)$. We aim to train a neural network, $s_{1}: \mathcal{M}^{2} \times \mathbb{R}_{+} \rightarrow T\mathcal{M}$ with parameters $\theta$ to approximate $\nabla_{y} \log p_{t}(x,y)$, by minimizing
\begin{equation}
    \mathbb{E}_{\mathbb{P}_{t}(x,y)}\left[|| \nabla_{y} \log p_{t}(x,y)-s_{1}(x,y,t;\theta)||_{2}^{2}\right],
    \label{eq:score_loss_fischer}
\end{equation}
where $\mathbb{P}_{t}(x,y)$ denotes the law of Brownian motion on $\mathcal{M}$. \citet{debortoli2022riemannian} shows that $\ell_{t}$ can be minimized implicitly without evaluating $\log p_{t}(x,y)$ using the loss function
\begin{equation}
    \mathbb{E}_{\mathbb{P}_{t}(x,y)}\left[\norm{s_{1}(x,y,t;\theta)}_{2}^{2}+2\normalfont{\text{div}}(s_{1}(x,\cdot,t;\theta))(y)\right],
    \label{eq:score_loss_fun}
\end{equation}
From eq.~\ref{eq:score_loss_fun} it can be seen that we can minimize $\ell_{t}$ without knowing $\log p_{t}(x,y)$. To avoid evaluating the computational expensive divergence in eq.~\ref{eq:score_loss_fun}, we will estimate the score by denoising score matching (DSM) \citep{pascal_dsm}
\begin{equation}
    \frac{1}{2}\mathbb{E}_{\mathbb{P}_{t}(x,y)}\left[\norm{s_{1}(x,y,t;\theta)+\frac{z}{\sigma}}_{2}^{2}\right],
    \label{eq:score_loss_dsm}
\end{equation}
where $\sigma^{2}$ denotes the variance of the infinitesimal time step of Brownian motion and $z:=\frac{x-y}{\sigma}$. Note that eq.~\ref{eq:score_loss_dsm} holds in Euclidean cases. However, since Brownian motion on Riemannian manifolds locally behaves like Euclidean Brownian motion \citep{hsustochastic}, we can use eq.~\ref{eq:score_loss_dsm} to estimate the score as long as the time step if sufficiently small. Note that there are other loss functions such as sliced score matching \citep{song2019sliced} and variance reducing method \citep{wang2020wasserstein, meng2021estimating} that can similarly be applied. Estimating the above loss function requires only sampling paths of Riemannian Brownian motion, which can be done in either local coordinates \citep{hsustochastic} or using geodesic random walk in the tangent space \citep{jax2018github}. We refer to appendix~\ref{ap:sampling} for details on the sampling method.
\paragraph{Estimating the time derivative} Using the score above we are able to estimate the diffusion mean by gradient descent. To also estimate the diffusion variance, we use the following result in the proof of theorem~4.8 in \citet{eltzner2022diffusion} 
\begin{equation}
    \frac{\dif }{\dif t}\mathbb{E}\left[\log p_{t}(X,y)\right] = \mathbb{E}\left[\frac{1}{2} \left(\Delta_{y}\log p_{t}(X,y) + ||\nabla_{y} \log p_{t}(X,y)||^{2}\right)\right]
    \label{eq:gradt_log}
\end{equation}
where $\Delta_{y}\log p_{t}(x,y) = g^{jk}\frac{\partial^{2} \log p}{\partial y^{j}\partial y^{k}}-g^{jk}\Gamma_{jk}^{l}\frac{\partial \log p}{\partial y^{l}}$. Thus, the time gradient can be estimated using the score and the trace of second order score. \citep{meng2021estimating} proposes methods to estimate higher-order scores, but we found empirically that it was more stable using automatic differentiation of the score to estimate the trace of the Hessian of the heat kernel.
\clearpage
\begin{wrapfigure}{r}{0.5\textwidth}
    \centering
    \includegraphics[width=0.50\textwidth]{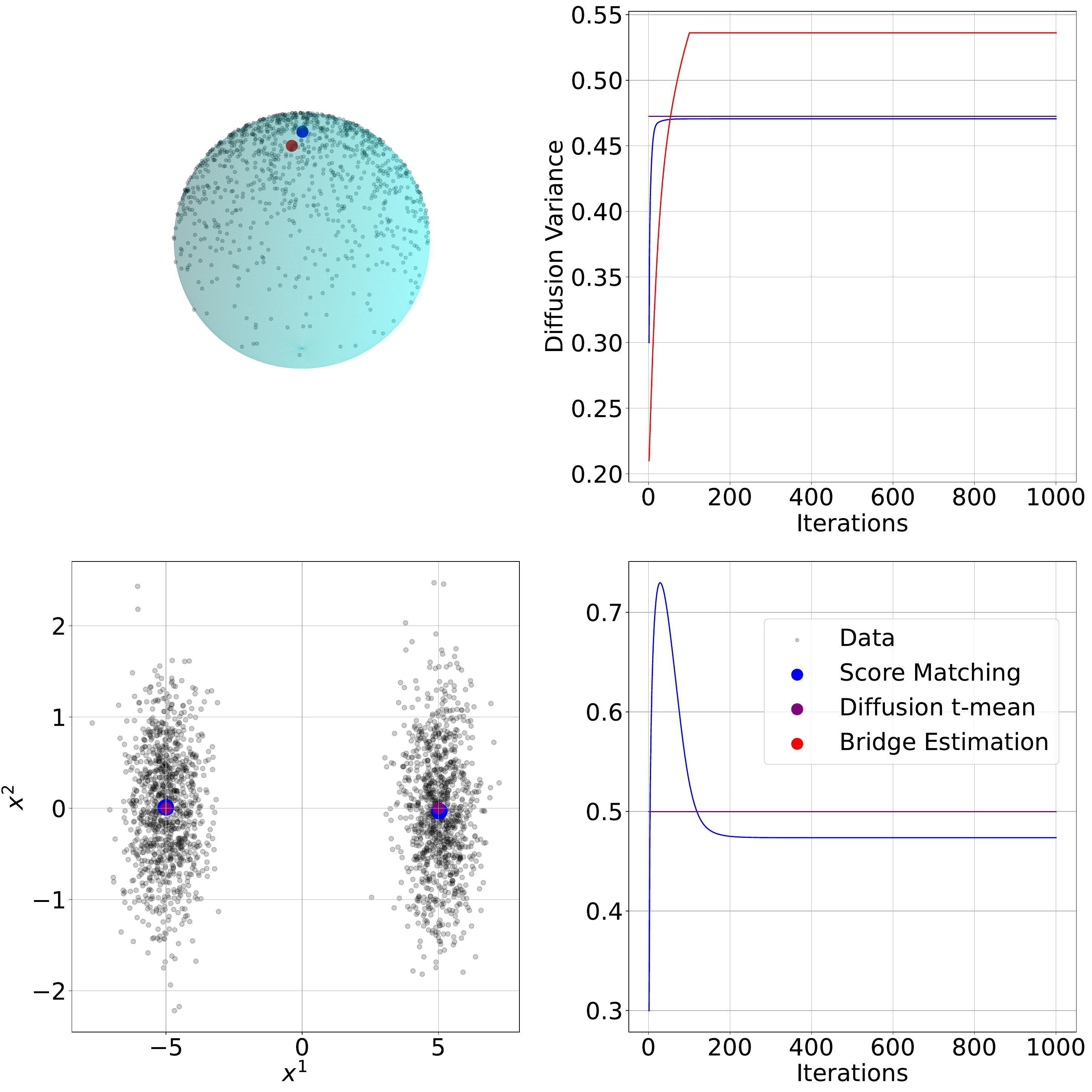}
    \caption{Estimates of the diffusion t-mean for $\mathbb{S}^{2}$ (first row) and $2 \times 2$ landmark shape space (second row) for synthetic data corresponding to $1,000$ samples of Brownian motion at time $t=0.5$. The first column shows the diffusion mean, while the second column shows the estimated diffusion variance.}
    \label{fig:score_synthetic_data}
    \vspace{-3.0em}
\end{wrapfigure}
\paragraph{Complexity in the manifold dimension.}
Numerically integrating geodesics using eq.~\ref{eq:geodesic_ode} requires inversion of the metric tensor in the Christoffel symbols, which scales cubicly in the manifold's dimension. Similarly, sampling Brownian motions requires finding the square root of the inverse metric, again with cubic scaling in the manifold dimension. Thus, finding the Fr\'echet mean via iterative optimization and diffusion means via bridge sampling has cubic scaling. For the proposed scheme, the cubic scaling persists for the training phase, but, for fixed network layout, inference time complexity scales only linearly in the input dimension.
\paragraph{Implementation.} In algorithm \ref{al:est_diffusion_mean}, the step on the Riemannian manifold is done using the exponential map. In practice, the step can also be done without the exponential map in the chart of the Riemannian manifold. This can be more efficient, especially for learned Riemannian manifolds. Further, it should be noted that the estimation of the diffusion $t$-mean using gradient descent will have local convergence, potentially not finding the entire set of diffusion $t$-means. We will restrict the diffusion variance $t \in (0,1)$ corresponding to the values of $t$ used for learning the score and time derivative. We note also that since the score is an approximation of the gradient of the log heat kernel, then convergence is not necessarily guaranteed. However, there exists well-known bounds on the score \citep{debortoli2022riemannian, song2021maximum}, we refer to these for details. We estimate the bridge estimate of the diffusion mean in accordance with \citet{jensen2023simulation}, which requires evaluating the logarithmic map. We note that there other method such as \citet{corstanje2024simulating}, which can be used if there is a closed-form expression of the heat kernel on a comparison manifold.

%% file: Sections/experiments.tex
The following section contains experiments and results regarding the estimation of the diffusion t-mean. Further we show its applicability to estimating the Fréchet mean as well as statistical algorithms on Riemannian manifolds. A full description of the manifolds used in the section can be found in appendix~\ref{ap:manifolds}. Details on implementation and training is described in appendix~\ref{ap:implementation}.
\paragraph{Estimating the Diffusion t-Mean.} To illustrate the applicability of using scores we compute the diffusion mean of a high-dimensional $30 \times 2$ landmarks shape space \citep{shapes_chapter}. We compute the diffusion mean on landmarks of butterfly wings as well on cardiac data, where bridge sampling is too expensive to compute. We also compute the diffusion mean on the expected metric for a learned manifold using a Gaussian process ($\mathcal{GP}$) \citep{tosi2014metrics} applied to a rotated MNIST image \citep{deng2012mnist}, where no closed form expressions are available. The estimated diffusion means can be seen in Fig.~\ref{fig:score_real_data}, where we see that the estimated diffusion means seem reasonable given the observed data. Table \ref{tab:diffusion_mean} shows the estimated diffusion mean using scores compared to bridge sampling for different manifolds and varying dimension, where we plot the result for $\mathbb{S}^{d}$ in Fig.~\ref{fig:timing_Sphere}. Each manifold has a sample of $1,000$ end points of Brownian motion with a fixed starting point and fixed diffusion time, $t=0.5$. Fig.~\ref{fig:score_synthetic_data} shows the estimates and sampled data for $\mathbb{S}^{2}$ and $2 \times 2$ landmarks. When the logarithmic map is available in closed-form, we compute the diffusion mean using bridge sampling. If the heat kernel is not known, we compare the estimates to the starting point and diffusion variance of the sampled data. From table \ref{tab:diffusion_mean} we see that using the score the estimates of the diffusion t-mean is comparable to bridge sampling, while having significantly lower computational time.
\begin{table*}[t]
    \centering
    \scriptsize
    \begin{tabular}{p{1.0cm} | c c | c c | c r}
    \textbf{Manifold} & $\pmb{||\mu^{\text{score}}-\mu_{0}||_{2}}$ & $\pmb{||\mu^{\text{bridge}}-\mu_{0}||_{2}}$ & $\pmb{|t^{\text{score}}-t_{0}|}$ & $\pmb{|t^{\text{bridge}}-t_{0}|}$ & \textbf{Score Time (s)} & \textbf{Bridge Time (s)} \\
    \hline
    $\mathbb{R}^{2}$ & $0.0172$ & $\pmb{0.0168}$ & $\pmb{0.0120}$ & $0.0406$ & $\pmb{0.3324 \pm 0.0127}$ & $0.3867 \pm 0.0036$ \\
    $\mathbb{R}^{3}$ & $\pmb{0.0269}$ & $0.1027$ & $\pmb{0.0092}$ & $0.0603$ & $\pmb{0.3449 \pm 0.0063}$ & $0.4505 \pm 0.0064$ \\
    $\mathbb{R}^{5}$ & $\pmb{0.0400}$ & $0.0779$ & $\pmb{0.0001}$ & $0.0585$ & $\pmb{0.3668 \pm 0.0084}$ & $0.6140 \pm 0.0086$ \\
    $\mathbb{R}^{10}$ & $\pmb{0.0395}$ & $0.1055$ & $\pmb{0.0079}$ & $0.0564$ & $\pmb{0.4341 \pm 0.0241}$ & $1.0747 \pm 0.0190$ \\
    \hline
    $\mathbb{S}^{2}$ & $\pmb{0.0022}$ & $0.1749$ & $\pmb{0.0096}$ & $0.0148$ & $\pmb{1.1519 \pm 0.0130}$ & $3.8635 \pm 0.0203$ \\
    $\mathbb{S}^{3}$ & $\pmb{0.0027}$ & $0.1072$ & $\pmb{0.0060}$ & $0.0371$ & $\pmb{1.1534 \pm 0.0205}$ & $12.2643 \pm 0.0878$ \\
    $\mathbb{S}^{5}$ & $\pmb{0.0107}$ & $0.1777$ & $\pmb{0.0074}$ & $0.0287$ & $\pmb{1.3828 \pm 0.0178}$ & $24.7303 \pm 0.0462$ \\
    $\mathbb{S}^{10}$ & $\pmb{0.0771}$ & $0.6015$ & $\pmb{0.0198}$ & $0.0241$ & $\pmb{1.8075 \pm 0.0240}$ & $91.7757 \pm 0.8074$ \\
    \hline
    $\mathrm{Sym}(2)$ & $\pmb{0.0584}$ & $1.6426$ & $\pmb{0.0337}$ & $0.0813$ & $0.8898 \pm 0.0111$ & $\pmb{0.6773 \pm 0.0061}$ \\
    $\mathrm{Sym}(3)$ & $\pmb{0.0336}$ & $2.4352$ & $0.0977$ & $\pmb{0.0876}$ & $\pmb{1.0214 \pm 0.0097}$ & $1.1938 \pm 0.0424$ \\
    $\mathrm{Sym}(5)$ & $\pmb{0.0653}$ & $3.8941$ & $0.1202$ & $\pmb{0.0905}$ & $\pmb{1.4501 \pm 0.0102}$ & $2.6307 \pm 0.1142$ \\
    \hline
    $\mathrm{LM}(2\times 2)$ & $0.0473$ & $-$ & $0.0263$ & $-$ & $1.3179 \pm 0.0119$ & $-$ \\
    $\mathrm{LM}(3\times 2)$ & $0.0384$ & $-$ & $0.0118$ & $-$ & $1.4223 \pm 0.0159$ & $-$ \\
    $\mathrm{LM}(5\times 2)$ & $0.0704$ & $-$ & $0.0212$ & $-$ & $1.6141 \pm 0.0163$ & $-$ \\
    \hline
    \end{tabular}
    \caption{The estimated values and runtime of the diffusion $t$-mean using score and bridge sampling for different manifolds and dimension. The data consists of $1,000$ paths of Riemannian Brownian motion at time $T=0.5$. For $\mathbb{S}^{10}$ we only use $100$ samples. If the heat kernel is not known in closed-form we compare the estimates using scores and bridge sampling to the staring point and diffusion variance $T=0.5$. For details on the heat kernels, see appendix~\ref{ap:heat_kernel}. We only compute the diffusion mean with bridge sampling when the logarithmic map is available in closed-form. The estimated run-time is based on five repeats after an initial compile of the code and show the time for $5$ iterations. For details on hardware we refer to appendix~\ref{ap:implementation}. LM referes to Landmarks.}
    \label{tab:diffusion_mean}
\end{table*}
\begin{table*}[!b]
    \centering
    \scriptsize
    \begin{tabular}{c | c c | c r}
    \textbf{Manifold} & $\pmb{||\mu^{\text{score}}-\mu_{0}||_{2}}$ & $\pmb{||\mu^{\text{Fr\^echet}}-\mu_{0}||_{2}}$ & \textbf{Score Time (s)} & \textbf{Fr\^echet Time (s)} \\
    \hline
    $\mathbb{R}^{2}$ & $\pmb{0.0206}$ & $0.0348$ & $0.0869 \pm 0.0020$ & $\pmb{0.0407 \pm 0.0407}$ \\
    $\mathbb{R}^{3}$ & $\pmb{0.0199}$ & $0.0664$ & $0.0910 \pm 0.0011$ & $\pmb{0.0399 \pm 0.0399}$ \\
    $\mathbb{R}^{5}$ & $\pmb{0.0214}$ & $0.0638$ & $0.0877 \pm 0.0017$ & $\pmb{0.0420 \pm 0.0420}$ \\
    $\mathbb{R}^{10}$ & $\pmb{0.0161}$ & $0.0600$ & $0.0952 \pm 0.0025$ & $\pmb{0.0487 \pm 0.0487}$ \\
    \hline
    $\mathcal{S}^{2}$ & $\pmb{0.0661}$ & $0.1352$ & $0.5629 \pm 0.0225$ & $\pmb{0.3021 \pm 0.3021}$ \\
    $\mathbb{S}^{3}$ & $\pmb{0.0477}$ & $0.3562$ & $0.4425 \pm 0.0055$ & $\pmb{0.2356 \pm 0.2356}$ \\
    $\mathbb{S}^{5}$ & $\pmb{0.0536}$ & $1.0873$ & $0.4873 \pm 0.0268$ & $\pmb{0.2623 \pm 0.2623}$ \\
    $\mathbb{S}^{10}$ & $\pmb{0.5727}$ & $1.5465$ & $0.4993 \pm 0.0108$ & $\pmb{0.2768 \pm 0.2768}$ \\
    \hline
    $\mathcal{P}(2)$ & $0.0134$ & $\pmb{0.0000}$ & $0.6206 \pm 0.0102$ & $\pmb{0.6399 \pm 0.6399}$ \\
    $\mathcal{P}(3)$ & $0.0085$ & $\pmb{0.0000}$ & $\pmb{0.6534 \pm 0.0097}$ & $0.7171 \pm 0.7171$ \\
    $\mathcal{P}(5)$ & $0.0204$ & $\pmb{0.0000}$ & $\pmb{0.7284 \pm 0.0089}$ & $0.7767 \pm 0.7767$ \\
    \hline
    $\mathrm{LM}(2\times 2)$ & $0.0532$ & $-$ & $0.3305 \pm 0.0060$ & $-$ \\
    $\mathrm{LM}(3\times 2)$ & $0.0303$ & $-$ & $0.3511 \pm 0.0032$ & $-$ \\
    $\mathrm{LM}(5\times 2)$ & $0.0762$ & $-$ & $0.3671 \pm 0.0042$ & $-$ \\
    \hline
    \end{tabular}
    \caption{The estimated values and runtime of the Fr\^echet mean using the score and bridge sampling for different manifolds and dimension. The data is the same as in table~\ref{tab:diffusion_mean}. We only compute the Fr\^echet mean with the logarithmic map when the logarithmic map is available in closed-form. Expect the Eucldiean case we compare it to the starting point of the data sampled. The estimated run-time is based on five repeats after an initial compile of the code and show the time for $5$ iterations. For details on hardware we refer to appendix~\ref{ap:implementation}.}
    \label{tab:frechet_mean}
\end{table*}
\paragraph{Estimating the Logarithmic map and Fréchet mean} As a side-effect of computing the score, we estimate the Logarithmic map as the time-limit of the score.  \citet{hsustochastic} shows that $\lim_{t \searrow 0} t \log p(x,y,t) = -\frac{\dist^{2}\left(x,y\right)}{2}$ uniformly, which implies that
\begin{equation}
    -\lim_{t \searrow 0} t \nabla_{y}\log p(x,y,t) = \text{Log}_{y}(x).
    \label{eq:learned_log}
\end{equation}
Thus, choosing a sufficiently small $T>0$, we can approximate the Logarithmic map as the limit of the learned gradients. Further, the Fréchet mean is given by $\mu^{\text{Fréchet}}(x_{1:N}) = \argmin_{\mu \in \mathcal{M}} \mathbb{E}\left[\dist^{2}(X,\mu)\right]$ \citep{frechet1948}, which can be estimated by gradient descent with $\nabla_{\mu}\mathbb{E}\left[\dist^{2}(X,\mu)\right] = \mathbb{E}\left[2\text{Log}_{\mu}(X)\right]$ \citep{pennec2006statriemann}. Choosing a sufficiently small $T>0$, we can approximate the Logarithmic map as the limit of the learned gradients using eq.~\ref{eq:learned_log} with gradient-descent. Since we only have to evaluate the scores in a fixed, sufficiently small value $T>0$, we train specifically the scores for this approximation by only evaluating at a fixed time to increase performance. We set $t=0.01$ for these experiments. Similar to \citet{geodesics_heat}, we normalize in the gradient steps the estimated Logarithm map in eq.~\ref{eq:learned_log}, since only the direction of the Logarithm map is important in the gradient descent method. We illustrate the estimate on DTI tensor data \citep{dti_ref1, dti_ref2, dti_ref3} in Fig.~\ref{fig:score_frechet_spdn}. The DTI tensor data consists of coronal slice of an HCP subject, which corresponds to $3 \times 3$ positive definite symmetric matrices. We plot the two leading eigenvalues of the data as well as the estimate using the closed-form logarithmic map compared to the logarithmic map estimated by the scores. Table~\ref{tab:frechet_mean} illustrates the estimates of the Fréchet mean using scores compared to using the closed-form expression of the logarithmic map. From Table~\ref{tab:frechet_mean} we see that the difference between the estimates is low, while the computational time is also reasonable. Note that the computational time of ground truth is significantly faster, since we use closed form expressions of the logarithmic map rather than solving the \textsc{ODE} \eqref{eq:geodesic_ode} or minimizing the energy functional.

\begin{wrapfigure}{R}{0.40\textwidth}
    \vspace{-0.0em}
    \centering
    \includegraphics[width=0.40\textwidth]{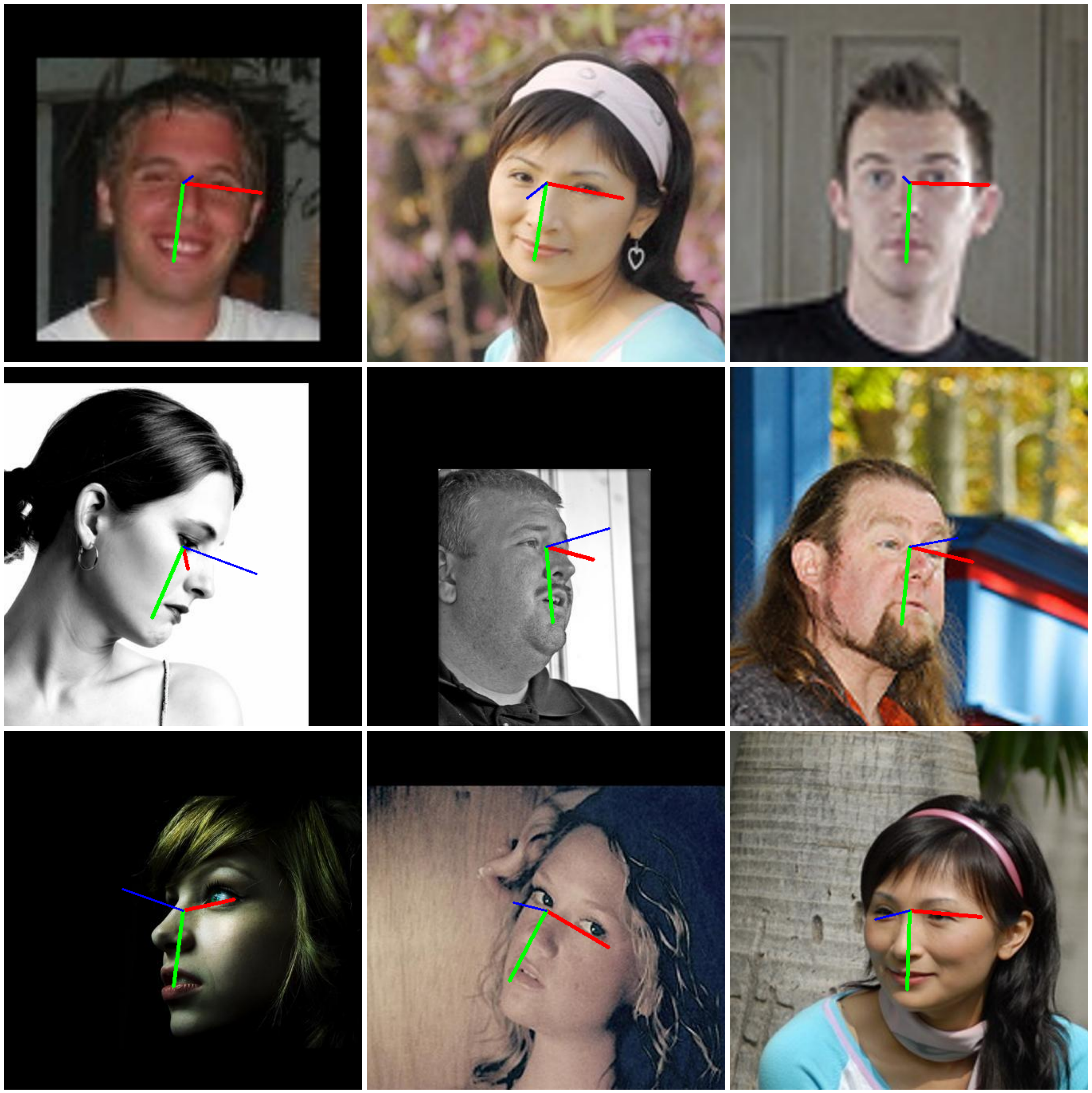}
    \caption{Each row shows the three images from the same cluster using Riemannian $K$-means algorithm on the AFLW-2000 dataset. The Riemannian $K$-means algorithm is estimated for the corresponding points of the head position in $\mathbb{S}^{3}$.}
    \label{fig:rm_kmeans_head_pose}
    \vspace{-5.5em}
\end{wrapfigure}
\paragraph{Riemannian $k$-means} The $k$-means algorithm is an unsupervised learning \citep{bishop}, which has been generalized into a Riemmannian setting \citep{arvanitidis2018latent}. It consists of initializing K centroids and then assign each data point to the cluster of the closest centroid, and then iteratively update the centroid of each cluster as the mean value of data points in the cluster to minimize $\sum_{i=1}^{N}\sum_{k=1}^{K} \delta_{ik}||x_{i}-\mu_{k}||^{2}$ with $x_{1:N}$ denoting the observations and $\delta_{ik} \in \{0,1\}$ for all $i,k$ with $\sum_{k}\delta_{ik}=1$. Similarly to the Euclidean method we can update the centroids computing the Fréchet mean using the gradient descent using eq. \ref{eq:learned_log}. In order to assign observations to their respective cluster, we have to use to estimate the Riemannian distance. By Varadhan's formulas \citep{varadhan_heat} we have that $\dist(x,y)^{2} = -4 \lim_{t\searrow 0} t\log p_{t}(x,y)$. It follows from l'H\^ospitals rule that
\begin{equation}
    \begin{split}
        \lim_{t\searrow 0} t \log p_{t}(x,y) &= \lim_{t\searrow 0}\frac{\log p_{t}(x,y)}{1/t} \\
        &= \lim_{t\searrow 0} \frac{\partial_{t} \log p_{t}(x,y)}{\frac{\dif}{\dif t} \frac{1}{t}} \\
        &= \lim_{t \searrow 0} -\partial_{t} \log p_{t}(x,y)t^{2}.
    \end{split}
    \label{eq:learned_dist}
\end{equation}
As argued in \citep{geodesics_heat} Varadhan's formula is sensitive to approximation errors. However, since we in the $k$-means algorithm only use the distance for cluster assignment, we are not required to have an accurate approximation of the distance. The algorithm can be found in appendix~\ref{ap:implementation}. To illustrate the method we consider the AFLW-2000 dataset \citep{aflw2000}, which consists of 1667 images with landmarks annotation as well Euler angles corresponding to the head pose of face images. The Euler angles can be converted into points on $\mathbb{S}^{3}$. We compute the $K$-means estimate on $\mathbb{S}^{3}$ and show the corresponding images for $K=3$ in Fig.~\ref{fig:rm_kmeans_head_pose}. We see that the method seem to accurately distinguish the images between looking straight ahead, looking right and looking to the left.
%

%
\begin{wrapfigure}{r}{0.3\textwidth}
    \vspace{-2.0em}
    \centering
    \includegraphics[width=0.3\textwidth]{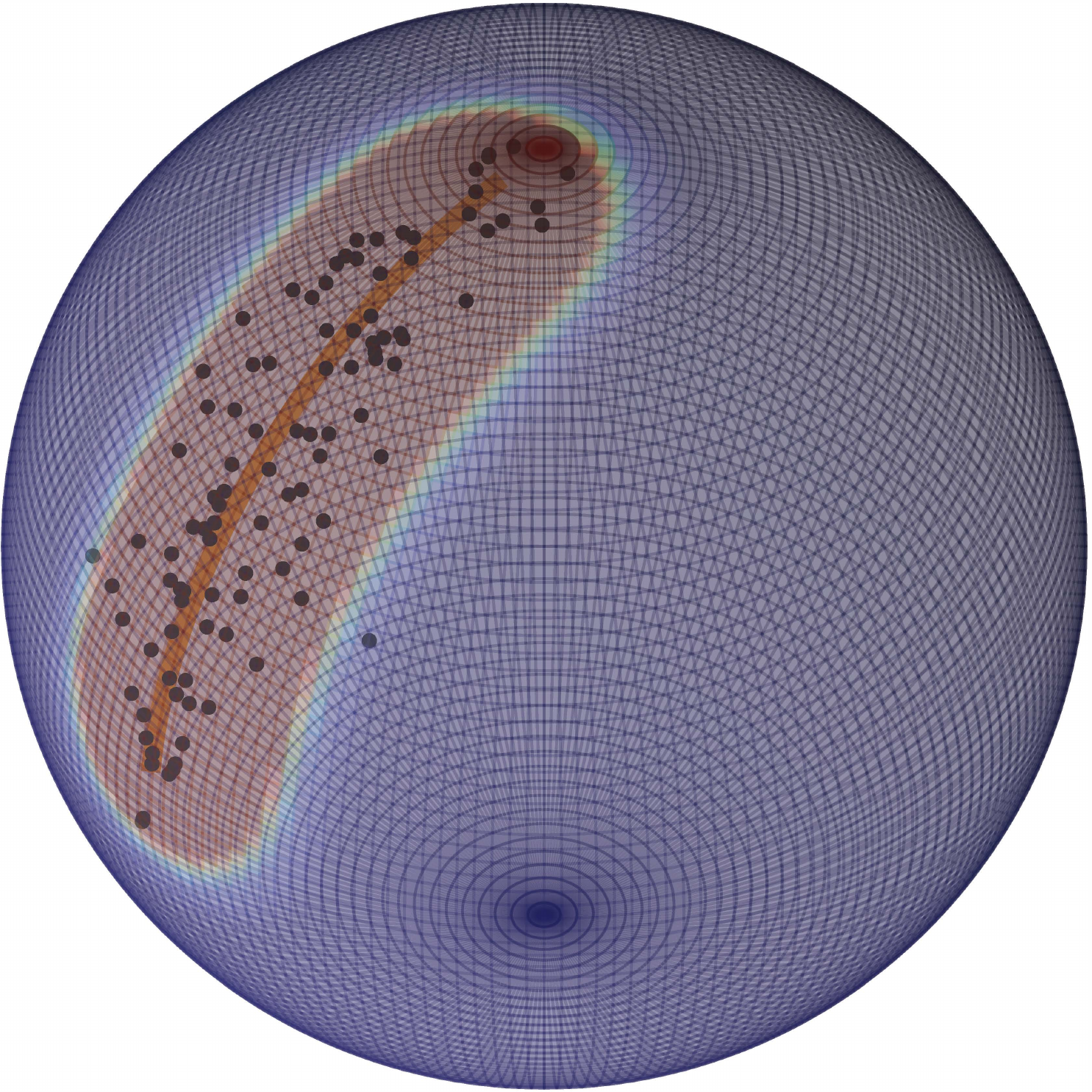}
    \caption{Maximum likelihood geodesic regression (orange) on $\mathbb{S}^{2}$ for synthetic data corresponding to noisy measurements (black) around a geodesic.}
    \label{fig:mlgr_synthetic_sphere}
    \vspace{-2.0em}
\end{wrapfigure}
\paragraph{Maximum Likelihood Riemannian Regression (MLRR)} Geodesic regression (GR) aims to find a relationship between observations $(x_{1},y_{1}), \dots, (x_{N},y_{N})$ by minimizing $\min_{\mu, v} \sum_{i=1}^{N}\dist\left(y_{i},\mathrm{Exp}\left(\mathrm{Exp}(\mu, x_{i}v), \epsilon_{i}\right)\right)^{2}$, where $\epsilon \in T\mathcal{M}$ \citep{fletcher:geodesicregression:2013} such that the observations $y_{i}$ is described by a geodesic at $\mu$ with initial velocity $x_{i}v \in T_{\mu}\mathcal{M}$. Instead of having the observation noise $\epsilon$ in the tangent bundle we will consider the measurement noise as independent dispersions of Riemannian Brownian motion similar to \citep{sommer2018infinitesimal}. Rather than restricting the regression to a geodesic we will consider a general function, $f_{\theta}: \mathbb{R}^{d} \rightarrow \mathcal{M}$, with parameters $\theta$, and a diffusion variance function, $\sigma_{\Phi}: \mathbb{R}^{d} \rightarrow \mathbb{R}$, with parameters $\Phi$ such that
\begin{equation*}
    y_{i} \sim W^{\mathcal{M}}_{\sigma_{\Phi}(x)^{2}}\left(f_{\theta}(x)\right),
\end{equation*}
such that the observations can be seen as samples around a diffusion mean along $f$. Unlike the original formulation in \citet{fletcher:geodesicregression:2013} the measurement noise is not restricted as geodesics from the tangent space in direction $\epsilon_{i}$, but rather as measurement noise directly on the manifold by Brownian motion. Assuming independence between the observations we have that the log likelihood is given by
\begin{equation*}
    \mathcal{L} = \sum_{i=1}^{N} \log p_{\sigma_{\Phi}(x)^{2}}\left(f_{\theta}(x), y_{i}\right),
\end{equation*}
where $\sigma^{2}$ corresponds to the diffusion variance and $p$ the heat kernel. By symmetry of heat kernel, i.e. $\nabla_{x}p_{t}(x,y)=\nabla_{y}p_{t}(x,y)$, then the gradient of the $\mathrm{log}$ likelihood is given by
\begin{equation}
    \begin{split}
        \nabla_{\theta}\mathcal{L} &= \sum_{i=1}^{N} \nabla_{y}\log p_{\sigma_{\Phi}(x)^{2}}(f_{\theta}(x), y_{i})\nabla_{\theta}f_{\theta}(x), \\
        \nabla_{\Phi}\mathcal{L} &= 2\sum_{i=1}^{N} \partial_{t} \log p_{\sigma_{\Phi}(x)^{2}}(f_{\theta}(x), y_{i})\nabla_{\Phi}\sigma_{\Phi}(x),
    \end{split}
\end{equation}
We therefore propose to estimate the parameters by gradient descent. If $\sigma$ is constant and $f_{\theta} = \mathrm{Exp}$, then the above formulation corresponds to maximum likelihood geodesic regression. Note further that if $\sigma\rightarrow 0$ if fixed, then the objective function above is equivalent to minimizing the distance by \citet{hsustochastic}. In this case the original objective function from \citet{fletcher:geodesicregression:2013} appears. We show in Fig.~\ref{fig:mlgr_synthetic_sphere} the result of maximum likelihood geodesic regression on $\mathbb{S}^{2}$. Further using the AFLW-2000 dataset we estimate the head position given landmark observations on the images. We show the result in Fig. \ref{fig:head_pose_s3}. In order to compute the uncertainty estimate on $\mathbb{S}^{3}$ we train a neural network $g_{t}(x,y)$ such that $\nabla_{y} g_{t}(x,y)$ approximates the score as described in \citet{learn_log_p}.

%% file: General/conclusion.tex
In this paper we have shown that the Riemannian diffusion mean can be estimated using scores from diffusion models. The estimation of the diffusion mean was empirically compared to bridge simulation to estimate the diffusion mean, where score matching performed well in terms of both the estimations error and computational time. We further showed that the method extends to computing the Fr\^echet mean efficiently. We showed the applicability of the method by introducing a Riemannian $K$-means algorithm as well as maximum likelihood Riemannian regression model, where distances or the density are not known in closed-form expression. This work shows how the score of Riemannian Brownian motion can be used to build efficient algorithms on general Riemannian manifolds.

\paragraph{Limitations and further research.} In the paper we have assumed that the manifolds are geodesically and stochastically complete, have bounded sectional curvature, orientable and boundary less. Even though this covers most manifolds some will not be able fall into this category. Further research can be done to generalize the diffusion mean and sampling of Brownian motion to other cases that can generalize the above approach. The training of the score function can be numerically expensive if no closed-form expressions exist, which can restrict learning the score for high-dimensional data. In this paper we only estimate the log gradient of the score, however, to improve the model higher order derivatives could also be investigated to make the optimization scheme for the diffusion mean more efficient. We also note that we learn the score by simulating paths on the manifold. However, for high dimensional spaces there might be regions, where the score has not seen sampled paths before, which would make the estimate of the score less accurate.

%% file: General/appendix/sampling.tex
Brownian motion, $W_{t}^{\mathcal{M}}$ on a Riemannian manifold, $\mathcal{M}$, can be simulated as a random walk in the tangent space mapped onto the Riemannian manifold by the exponential map \citep{Jrgensen1975TheCL}.

\begin{algorithm}
    \SetAlgoLined
    \textbf{Input}: $T$, $x_{0} \in \mathcal{M}$, $N_{\text{steps}}$, $D$ (dimension of manifold) \\
    \textbf{Output}: Samples of Brownian Motion $W_{t}$ with $W_{0}=x_{0}$ \\
    Set $\delta = \frac{T}{N_{\text{steps}}}$ \\
    Set $W_{0} = x_{0}$ \\
    \For{$k=1,\dots,N_{\text{steps}}$} {
        Sample $v_{k} \sim \mathcal{N}\left(0, I_{D}\right)$ \\
        $W_{k\delta} = \text{Exp}_{\sqrt{\delta}v}\left(W_{(k-1)\delta}\right)$
    }
    return $\{W_{t}\}_{t \in [0,\delta, 2\delta, \dots, T]}$ \\
    \caption{Simulating Brownian Motion in Tangent Space}
    \label{al:sampling_tangent_space}
\end{algorithm}

Brownian Motion on a manifold can equivalently also be simulated in a local coordinate system, $\{x^{i}\}_{i=1}^{D} \subset \mathcal{M}$ with $\sigma^{i}_{j}=g^{ji}$ \citep{hsustochastic}

\begin{algorithm}
    \SetAlgoLined
    \textbf{Input}: $T$, $x_{0} \in \mathcal{M}$, $N_{\text{steps}}$, $D$ (dimension of manifold) \\
    \textbf{Output}: Samples of Brownian Motion $W_{t}$ with $W_{0}=x_{0}$ \\
    Set $\delta = \frac{T}{N_{\text{steps}}}$ \\
    Set $W_{0} = x_{0}$ \\
    \For{$k=1,\dots,N_{\text{steps}}$} {
        Sample $v_{k} \sim \mathcal{N}\left(0, I_{D}\right)$ \\
        $W_{k\delta}^{i} = \frac{1}{2}g^{jk}\left(W_{(k-1)\delta}\right)\Gamma_{jk}^{i}\left(W_{(k-1)\delta}\right)dt+\sigma^{i}_{j}\left(W_{(k-1)\delta}\right)v_{k}^{j}, \quad \forall i \in \{1,\dots, D\}$
    }
    return $\{W_{t}\}_{t \in [0,\delta, 2\delta, \dots, T]}$ \\
    \caption{Simulating Brownian Motion in Local Coordinates}
    \label{al:sampling_local_coordinates}
\end{algorithm}

%% file: General/appendix/manifolds.tex
Table \ref{tab:manifold_description} describes the various manifolds used in the papers as well as their embeddings.

\begin{table*}[t]
    \centering
    \scriptsize
    \begin{tabular}{p{2cm} | p{3cm} | p{2cm} | p{2cm} | p{2cm}}
    \textbf{Manifold} & \textbf{Description} & \textbf{Parameters} & \textbf{Local Coordinates} & \textbf{Embedding} \\
    \hline
    $\mathbb{R}^{n}$ & The Euclidean $n$-dimensional vector space & - & Standard basis of $\mathbb{R}^{n}$ & - \\
    \hline
    $\mathbb{S}^{n}$ & The n-sphere, $\{x \in \mathbb{R}^{n+1} \;|\; ||x||_{2}=1\}$ & - & Stereographic coordinates & $15800 \pm 239$ \\
    \hline
    $\mathcal{P}(n)$ & The space of positive definite matrices & - & Local coordinates are given respect to the standard basis in $\mathbb{R}^{n(n+1)/2}$ & Embedded into $\mathbb{R}^{n^{2}}$ by the mapping $f(x) = l(x)l(x)^{T}$, where $l: \mathbb{R}^{n(n+1)/2} \rightarrow \mathbb{R}^{n \times n}$ maps $x$ into a lower triangle matrix consisting of the elements in $x$. \\
    \hline
    $\mathrm{Sym}(n)$ & The space of symmetric matrices & - & Local coordinates are given respect to the standard basis in $\mathbb{R}^{n(n+1)/2}$ & Embedded into $\mathbb{R}^{n^{2}}$ by converting the local coordinates into a symmetric triangular matrix. \\
    \hline
    $\mathrm{Landmarks}(n \times d)$ & The space of landmarks, $\{x_{k}\}_{k=1}^{n}$ is a shape space that consist of $n$ distinct points $x_{k} \in \mathbb{R}^{d}$. The landmark space can be equipped with a right-invariant Riemannian metric \citep{shapes_chapter}. & The Gaussian Kernel with parameter $\alpha=1$ & All landmarks are given with respect to the standard basis in $\mathbb{R}^{nd}$ & -  \\
    \hline
    \end{tabular}
    \caption{Description of the manifolds used in the paper.}
    \label{tab:manifold_description}
\end{table*}

To avoid singularities in the chart for the embedded manifolds the basis of the chart is updated, when the chart deviates from the initial center. This is for instance the case for stereographic projection, where the antipodal point is undefined.

%% file: General/appendix/heat_kernels.tex
The following section will briefly state the heat kernels that have closed form expressions for the Riemannian manifolds studied in this paper.

\subsection{Euclidean Space}
In $\mathbb{R}^{m}$ the heat kernel is given by \citep{eltzner2022diffusion}

\begin{equation}
    p(x,y,t) = \frac{1}{(2\pi t)^{m/2}}e^{-\frac{-||x-y||^{2}}{2t}},
\end{equation}

with $x,y \in \mathbb{R}^{m}$. This implies that

\begin{equation}
    \ln p(x,y,t) = -\frac{||x-y||^{2}}{2t}-\frac{m}{2}\ln\left(2\pi t\right).
\end{equation}

Thus

\begin{equation}
    \begin{split}
        \nabla_{y} \ln p(x,y,t) &= \frac{x-y}{t}, \\
        \frac{d}{dt} \ln p(x,y,t) &= \frac{||x-y||^{2}}{2t^{2}}-\frac{m}{2t}
    \end{split}
    \label{eq:euc_true_gradients}
\end{equation}

It can clearly be seen that the optimal diffusion mean is given by

\begin{equation*}
    \frac{1}{N}\sum_{i=1}^{N}\nabla_{\mu} \ln p(x_{i},\mu^{*},t) \Rightarrow \mu^{*}=\frac{1}{N}\sum_{i=1}^{N}x_{i},
\end{equation*}

which is independent of the diffusion time. The optimal diffusion time is then given by

\begin{equation*}
    \frac{1}{N}\sum_{i=1}^{N}\frac{d}{dt} \ln p(x_{i},\mu^{*},t^{*}) = \frac{\frac{1}{N}\sum_{i=1}^{N}||x_{i}-\mu^{*}||^{2}}{m}.
\end{equation*}

\subsection{Circle}
The heat kernel on $\mathbb{S}^{1}$ is given by \citep{eltzner2022diffusion}

\begin{equation}
    p(x,y,t) = \frac{1}{\sqrt{2\pi t}}\sum_{k \in \mathbb{Z}}e^{-\frac{-(x-y+2\pi k)^{2}}{2t}},
\end{equation}

with $x,y \in \mathbb{R}/2\pi \mathbb{Z}$, which implies that

\begin{equation}
    \ln p(x,y,t) = -\frac{1}{2}\ln 2\pi t + \ln\sum_{k \in \mathbb{Z}}e^{-\frac{-(x-y+2\pi k)^{2}}{2t}}.
\end{equation}
 
Assuming that the infinite sum is uniformly and absolute convergent, then

\begin{equation}
    \begin{split}
        &\nabla_{y} \ln p(x,y,t) = \frac{\sum_{k \in \mathbb{Z}}\left(2\pi k + x -y\right)e^{-\frac{-(x-y+2\pi k)^{2}}{2t}}}{\sqrt{2\pi t}p(x,y,t)}, \\
        &\frac{d}{dt} \ln p(x,y,t) = \frac{\frac{1}{2\sqrt{2\pi t}}\sum_{k \in \mathbb{Z}}\frac{-(x-y+2\pi k)^{2}}{2t^{-2}}e^{-\frac{-(x-y+2\pi k)^{2}}{2t}}}{p(x,y,t)} - \frac{\frac{1}{(2t)^{3/2}\sqrt{\pi}}\sum_{k \in \mathbb{Z}}e^{-\frac{-(x-y+2\pi k)^{2}}{2t}}}{p(x,y,t)}
    \end{split}
    \label{eq:circle_true_gradients}
\end{equation}

\subsection{m-Sphere} 
For the m-sphere, $\mathbb{S}^{m}$, with $m \geq 2$ the heat kernel is given by uniformly and absolutely convergent infinite sum in theorem 1 in \cite{zhaohksm}

\begin{equation}
    p(x,y,t) = \sum_{l=0}^{\infty} e^{-l(l+m-1)\frac{t}{2}}\frac{2l+m-1}{m-1}\frac{1}{A_{S}^{m}}C_{l}^{(m-1)/2}\left(\langle x, y \rangle_{\mathbb{R}^{m+1}}\right),
\end{equation}

where $x,y \in \mathbb{R}^{m+1}$, $C_{l}^{\alpha}$ denotes the Gegenbauer polynomials and

\begin{equation}
    A_{S}^{m} = \frac{2\pi^{(m+1)/2}}{\Gamma\left((m+1)/2\right)},
\end{equation}

which gives the gradients

\begin{equation}
    \begin{split}
         \nabla_{y} \ln p(x, y, t) &= \frac{\sum_{l=0}^{\infty} e^{-l(l+m-1)\frac{t}{2}}(2l+m-1)\frac{1}{A_{S}^{m}}C_{l-1}^{(m+1)/2}\left(\langle x, y \rangle_{\mathbb{R}^{m+1}}\right)x}{p(x,y,t)}, \\
         \frac{d}{dt} \ln p(x, y, t) &= \frac{-\sum_{l=0}^{\infty} \frac{l(l+m-1)}{2}e^{-l(l+m-1)\frac{t}{2}}\frac{2l+m-1}{m-1}\frac{1}{A_{S}^{m}}C_{l}^{(m-1)/2}\left(\langle x, y \rangle_{\mathbb{R}^{m+1}}\right)}{p(x,y,t)}.
    \end{split}
\end{equation}

The above is implemented using the recursion

\begin{equation}
    l C_{l}^{(\alpha)}(x) = 2(l-1+\alpha)xC_{l-1}^{(\alpha)}-(n+2\alpha-2)C_{l-2}^{(\alpha)}(x).
\end{equation}

%% file: General/appendix/data.tex
The synthetic data consists of the end points of $1,000$ sampled Brownian motions using algorithm \ref{al:sampling_local_coordinates} with starting point, $x_{0}$, described in table \ref{tab:score_matching_architecture}. The cardiac data consist of annotated images of cardiac, while the butterfly data consists of annotated images of butterflies. We have scaled the cardiac data, so it has similar numerical values to the butterfly data to be in area, where the estimated score has seen sufficiently many sample paths. The DTI tensor data consists of pre-processed diffusion data of 20 subjects from the Q3 release of the Human Connectome Project (HCP) \citep{dti_data}. The data is processed in accordance with \citet{dti_data}, which gives corresponding values in $\mathcal{P}(3)$ for four segmentation. We use $100$ samples of segmentation 1 in examples. The MNIST data is described in \citet{deng2012mnist}. 

\begin{table*}[t]
    \centering
    \scriptsize
    \begin{tabular}{p{2cm} | p{1.7cm} | p{4cm} | p{2cm} | p{1.4cm}}
    \textbf{Manifold} & $\mathbf{x}_{0}$ & \textbf{Architecture} & \textbf{Activation Function} & \textbf{Embedded} \\
    \hline
    $\mathbb{R}^{n}$ & $\mathbf{0}_{n}$ & $\mathrm{Linear}(128,128,128)$ & \textsc{tanh} & - \\
    \hline
    $\mathbb{S}^{n}$ & North pole & $\mathrm{Linear}(512,512,512,512,512)$ & \textsc{tanh} & $\times$ \\
    \hline
    $\mathcal{P}(n)$ & $10\cdot\mathrm{diag}(n)$ & $\mathrm{Linear}(512,512,512)$ & \textsc{tanh} & $-$ \\
    \hline
    $\mathrm{Sym}(n)$ & $\mathrm{diag}(n)$ & $\mathrm{Linear}(512,512,512)$ & \textsc{tanh} & $-$ \\
    \hline
    $\mathrm{Landmarks}(n\times 2)$ & See description above & $\mathrm{Linear}(512,512,512)$ & \textsc{tanh} & $-$ \\
    \hline
    \end{tabular}
    \caption{The Architecture for learning the score}
    \label{tab:score_matching_architecture}
\end{table*}

%% file: General/appendix/implementation.tex
The implementation is based on \textsc{jaxgeometry}. \textsc{jaxgeometry} is open-source and can be found at \url{https://github.com/ComputationalEvolutionaryMorphometry/jaxgeometry}. The initial seed value is set to 2712 when generating random numbers.

\subsection{Hardware}
The score has all been trained for 24 hours on the system with $4$ nodes, where for each node is configured with
\begin{itemize}
    \item 2x Intel Xeon Processor 2650v4 (12 core, 2.20GHz)
    \item 256 GB memory
    \item FDR-Infiniband
    \item 480 GB-SSD disk
\end{itemize}

All other computations have been performed on a \textit{HP} computer with Intel Core i9-11950H 2.6 GHz 8C, 15.6'' FHD, 720P CAM, 32 GB (2$\times$16GB) DDR4 3200 So-Dimm, Nvidia Quadro TI2004GB Descrete Graphics, 1TB PCle NVMe SSD, backlit Keyboard, 150W PSU, 8cell, W11Home 64 Advanced, 4YR Onsite NBD.

\subsection{Sampling}
We sample paths of Brownian motion using algorithm \ref{al:sampling_local_coordinates} with $N_{\text{steps}}=1000$ and $T=1.0$. Initially, a batch Brownian motion paths are simulated from a fixed starting point, $x_{0} \in \mathcal{M}$. For each $x_{0}$ we sample a fixed number of paths, and similarly for each $t$ we sample a fixed number of times. For each batch of sample paths we choose \textsc{samples per $x_{0}$} end points of the samples as new starting points for the next batch of Brownian motion. Table \ref{tab:hyper_sampling} shows the hyper-parameters from sampling.

\begin{table}[ht]
    \centering
    \begin{tabular}{c| c}
        \hline
        \textbf{Parameter} & \textbf{Value} \\
        \hline
        Samples per $x_{0}$ & $1$ \\
        \hline
        Samples per $t$ & $100$ \\
        \hline
        Number of $x_{0}$ & $1024$ \\
        \hline
        Time Steps for sampling Brownian Motion & $100$ \\
        \hline
        Sampling time for Brownian Motion & $1.0$ \\
        \hline
    \end{tabular}
    \caption{Hyper-Parameters for Sampling}
    \label{tab:hyper_sampling}
\end{table}

\subsection{Score Matching}

Table~\ref{tab:score_matching_architecture} shows the architecture for each Riemannian manifold. The column, 'embedded', indicates whether the score was learned in the embedded space. If there is no mark, the score is learned in local coordinates. The column, 'Activation Function', indicates the activation function applied after each hidden layer. All scores are trained for 24 hours or up to a maximum of $50,000$ epochs.

The initial starting point, $x_{0}$, in table \ref{tab:score_matching_architecture} for the landmark spaces consisting of 2 and 5 landmarks are 2d points, where the first coordinate is equally spaces between $[-5,5]$, and the second coordinate is zero. For 10-landmarks and 20-landmarks $x_{0}$ is set as the first observation of the butterfly dataset with sub-sampling landmarks accordingly. In table \ref{tab:score_matching_architecture} the initial coordinate for SPDN is written in local coordinates.

All manifolds are trained using the ADAM optimizer with learning rate $0.001$ with linear annealing for the first $1000$ epcohs and the cosine annealing similar to \citet{debortoli2022riemannian}. For the embedded scores, the score is projected the tangent space in embedded space.

\subsection{Optimization for Diffusion $t$-Mean and Fréchet Mean}
The estimation of the diffusion $t$-mean is done using the gradient descent algorithm to jointly estimate $t$ and $\mu$. The hyper-parameters for estimating the diffusion t-mean is shown in table~\ref{tab:optimization_diffusion_mean}.

\begin{table*}[ht]
    \centering
    \scriptsize
    \begin{tabular}{p{2cm} | p{1.3cm} | p{0.25cm} | p{1.25cm} | p{2.25cm}}
    \textbf{Manifold} & \textbf{Learning rate} & $\pmb{t}_{0}$ & \pmb{Iterations} & \pmb{Method} \\
    \hline
    $\mathbb{R}^{n}$ & $0.1$ & $0.2$ & $1000$ & ADAM \\
    \hline
    $\mathbb{S}^{n}$ & $0.1$  & $0.2$ & $1000$ & Gradient-Descent \\
    \hline
    $\mathrm{Sym}(n)$ & $0.1$  & $0.2$ & $1000$ & ADAM \\
    \hline
    $\mathrm{Landmarks}(n)$ & $0.1$  & $0.2$ & $1000$ & ADAM \\
    \hline
    \end{tabular}
    \caption{The Hyper-parameters for estimating the diffusion t-mean.}
    \label{tab:optimization_diffusion_mean}
\end{table*}
The hyper-parameters for estimating the Fr\^echet mean is shown in table~\ref{tab:optimization_frechet_mean}
\begin{table*}[ht]
    \centering
    \scriptsize
    \begin{tabular}{p{2cm} | p{1.3cm} | p{0.25cm} | p{1.25cm} | p{2.25cm}}
    \textbf{Manifold} & \textbf{Learning rate} & $\pmb{t}_{0}$ & \pmb{Iterations} & \pmb{Method} \\
    \hline
    $\mathbb{R}^{n}$ & $0.1$ & $0.2$ & $1000$ & ADAM \\
    \hline
    $\mathbb{S}^{n}$ & $0.01$  & $0.2$ & $1000$ & Gradient-Descent \\
    \hline
    $\mathrm{Sym}(n)$ & $0.1$  & $0.2$ & $1000$ & ADAM \\
    \hline
    $\mathcal{P}(n)$ & $0.1$  & $0.2$ & $1000$ & ADAM \\
    \hline
    $\mathrm{Landmarks}(n)$ & $0.1$  & $0.2$ & $1000$ & ADAM \\
    \hline
    \end{tabular}
    \caption{The Hyper-parameters for estimating the Fr\^echet mean.}
    \label{tab:optimization_frechet_mean}
\end{table*}

\subsection{Riemannian K-Means Estimation}
We estimate the Logarithmic map with a fixed time of $0.1$. The Fréchet mean is estimated using $100$, a step size of $0.1$ and a fixed time of $1.0$. The $K$-means algorithm is estimated using $10$ iterations with $K=4$ for both $\mathbb{S}^{2}$ and 20-landmarks.
\begin{algorithm}[H]
    \SetAlgoLined
    \textbf{Input}: K, $N_{\text{iter}}$, $\{\mu_{i}^{\text{init}}\}_{i=1}^{K}$, $\{x_{n}\}_{i=1}^{N}$ \\
    \textbf{Output}: \textsc{K} Centroids and Clusters  \\
    Learn neural network $s_{\theta}$ minimizing eq. \ref{eq:score_loss_fun} \\
    Set $\{\mu_{i}\}_{i=1}^{K} = \{\mu_{(i)}^{\text{init}}\}_{i=1}^{K}$ \\
    \For{i=1,\dots,$N_{\text{iter}}$} {
        \For{j=1,\dots,$K$} {
            \For{n=1,\dots,$N$} {
                Set $d_{n} = \sqrt{4t^{2}\partial_{t}\log p_{t}(x_{n},\mu_{j})}$ for small $t$ \\
            }
            Set $z_{n,\argmin_{n} d_{n}}=1, z_{nk}=0, \forall k \neq \argmin_{n} d_{n}$ \\
        }
        Set $\mu_{k}$ as the Fréchet mean using eq. \ref{eq:learned_log} for all $\{x_{n}\}_{n=1}^{N}$ in cluster $k$.
    }
    return $(\{\mu_{k}\}_{k=1}^{K}, \{z_{nk}\}_{k=1}^{K}, \forall n \in \{1,\dots, N\})$  \\
    \caption{Riemannian $k$-Means}
    \label{al:rm_kmeans}
\end{algorithm}

\subsection{Gaussian Process}
We train Gaussian process for a MNIST image rotated $2\pi$, where we sample $200$ images along the rotation. We fix the latent variables as the $200$ corresponding points on the circle and compute the posterior distribution and expected Riemannian metric as described in \citet{tosi2014metrics}.

%% file: score_paper.bbl
\begin{thebibliography}{40}
\providecommand{\natexlab}[1]{#1}
\providecommand{\url}[1]{\texttt{#1}}
\expandafter\ifx\csname urlstyle\endcsname\relax
  \providecommand{\doi}[1]{doi: #1}\else
  \providecommand{\doi}{doi: \begingroup \urlstyle{rm}\Url}\fi

\bibitem[Arnaudon et~al.(2023)Arnaudon, Holm, and Sommer]{shapes_chapter}
Alexis Arnaudon, Darryl Holm, and Stefan Sommer.
\newblock \emph{Stochastic Shape Analysis}, pages 1325--1348.
\newblock Springer, Switzerland, 2023.
\newblock ISBN 9783030986605.
\newblock \doi{10.1007/978-3-030-98661-2_86}.
\newblock Publisher Copyright: {\textcopyright} Springer Nature Switzerland AG 2023.

\bibitem[Arvanitidis et~al.(2018)Arvanitidis, Hansen, and Hauberg]{arvanitidis2018latent}
Georgios Arvanitidis, Lars~Kai Hansen, and Søren Hauberg.
\newblock Latent space oddity: on the curvature of deep generative models.
\newblock In \emph{International Conference on Learning Representations}, 2018.
\newblock URL \url{https://openreview.net/forum?id=SJzRZ-WCZ}.

\bibitem[Bhattacharya and Patrangenaru(2003)]{bpc2003}
Rabi Bhattacharya and Vic Patrangenaru.
\newblock {Large sample theory of intrinsic and extrinsic sample means on manifolds}.
\newblock \emph{The Annals of Statistics}, 31\penalty0 (1):\penalty0 1 -- 29, 2003.
\newblock \doi{10.1214/aos/1046294456}.
\newblock URL \url{https://doi.org/10.1214/aos/1046294456}.

\bibitem[Bishop(2006)]{bishop}
Christopher~M. Bishop.
\newblock \emph{Pattern Recognition and Machine Learning (Information Science and Statistics)}.
\newblock Springer-Verlag, Berlin, Heidelberg, 2006.
\newblock ISBN 0387310738.

\bibitem[Bortoli et~al.(2022)Bortoli, Mathieu, Hutchinson, Thornton, Teh, and Doucet]{debortoli2022riemannian}
Valentin~De Bortoli, Emile Mathieu, Michael Hutchinson, James Thornton, Yee~Whye Teh, and Arnaud Doucet.
\newblock Riemannian score-based generative modelling, 2022.

\bibitem[Bradbury et~al.(2018)Bradbury, Frostig, Hawkins, Johnson, Leary, Maclaurin, Necula, Paszke, Vander{P}las, Wanderman-{M}ilne, and Zhang]{jax2018github}
James Bradbury, Roy Frostig, Peter Hawkins, Matthew~James Johnson, Chris Leary, Dougal Maclaurin, George Necula, Adam Paszke, Jake Vander{P}las, Skye Wanderman-{M}ilne, and Qiao Zhang.
\newblock {JAX}: composable transformations of {P}ython+{N}um{P}y programs, 2018.
\newblock URL \url{http://github.com/google/jax}.

\bibitem[Bui et~al.(2022)Bui, Pokern, and Dellaportas]{buiInferencePartiallyObserved2022}
Mai~Ngoc Bui, Yvo Pokern, and Petros Dellaportas.
\newblock Inference for partially observed {Riemannian} {Ornstein}-{Uhlenbeck} diffusions of covariance matrices, November 2022.
\newblock URL \url{http://arxiv.org/abs/2104.03193}.
\newblock arXiv:2104.03193 [stat].

\bibitem[Corstanje et~al.(2024)Corstanje, van~der Meulen, Schauer, and Sommer]{corstanje2024simulating}
Marc Corstanje, Frank van~der Meulen, Moritz Schauer, and Stefan Sommer.
\newblock Simulating conditioned diffusions on manifolds, 2024.

\bibitem[Crane et~al.(2013)Crane, Weischedel, and Wardetzky]{geodesics_heat}
Keenan Crane, Clarisse Weischedel, and Max Wardetzky.
\newblock Geodesics in heat: A new approach to computing distance based on heat flow.
\newblock \emph{ACM Transactions on Graphics}, 32\penalty0 (5):\penalty0 1–11, September 2013.
\newblock ISSN 1557-7368.
\newblock \doi{10.1145/2516971.2516977}.
\newblock URL \url{http://dx.doi.org/10.1145/2516971.2516977}.

\bibitem[Delyon and Hu(2006)]{DELYON20061660}
Bernard Delyon and Ying Hu.
\newblock Simulation of conditioned diffusion and application to parameter estimation.
\newblock \emph{Stochastic Processes and their Applications}, 116\penalty0 (11):\penalty0 1660--1675, 2006.
\newblock ISSN 0304-4149.
\newblock \doi{https://doi.org/10.1016/j.spa.2006.04.004}.
\newblock URL \url{https://www.sciencedirect.com/science/article/pii/S0304414906000469}.

\bibitem[Deng(2012)]{deng2012mnist}
Li~Deng.
\newblock The mnist database of handwritten digit images for machine learning research.
\newblock \emph{IEEE Signal Processing Magazine}, 29\penalty0 (6):\penalty0 141--142, 2012.

\bibitem[do~Carmo(1992)]{do1992riemannian}
M.P. do~Carmo.
\newblock \emph{Riemannian Geometry}.
\newblock Mathematics (Boston, Mass.). Birkh{\"a}user, 1992.
\newblock ISBN 9783764334901.
\newblock URL \url{https://books.google.dk/books?id=uXJQQgAACAAJ}.

\bibitem[Eltzner et~al.(2022)Eltzner, Hansen, Huckemann, and Sommer]{eltzner2022diffusion}
Benjamin Eltzner, Pernille Hansen, Stephan~F. Huckemann, and Stefan Sommer.
\newblock Diffusion means in geometric spaces.
\newblock \emph{Bernoulli (in print)}, 2022.

\bibitem[Fletcher(2013)]{fletcher:geodesicregression:2013}
P.~Thomas Fletcher.
\newblock Geodesic regression and the theory of least squares on riemannian manifolds.
\newblock \emph{International Journal of Computer Vision}, 105, 11 2013.
\newblock \doi{10.1007/s11263-012-0591-y}.

\bibitem[Fr\'echet(1948)]{frechet1948}
Maurice Fr\'echet.
\newblock Les \'el\'ements al\'eatoires de nature quelconque dans un espace distanci\'e.
\newblock \emph{Annales de l'institut Henri Poincar\'e}, 10\penalty0 (4):\penalty0 215--310, 1948.
\newblock URL \url{http://www.numdam.org/item/AIHP_1948__10_4_215_0/}.

\bibitem[Glasser et~al.(2013)Glasser, Sotiropoulos, Wilson, Coalson, Fischl, Andersson, Xu, Jbabdi, Webster, Polimeni, {Van Essen}, and Jenkinson]{dti_ref1}
Matthew~F. Glasser, Stamatios~N. Sotiropoulos, J.~Anthony Wilson, Timothy~S. Coalson, Bruce Fischl, Jesper~L. Andersson, Junqian Xu, Saad Jbabdi, Matthew Webster, Jonathan~R. Polimeni, David~C. {Van Essen}, and Mark Jenkinson.
\newblock The minimal preprocessing pipelines for the human connectome project.
\newblock \emph{NeuroImage}, 80:\penalty0 105--124, 2013.
\newblock ISSN 1053-8119.
\newblock \doi{https://doi.org/10.1016/j.neuroimage.2013.04.127}.
\newblock URL \url{https://www.sciencedirect.com/science/article/pii/S1053811913005053}.
\newblock Mapping the Connectome.

\bibitem[Hauberg et~al.(2015)Hauberg, Schober, Liptrot, Hennig, and Feragen]{dti_data}
S{\o}ren Hauberg, Michael Schober, Matthew Liptrot, Philipp Hennig, and Aasa Feragen.
\newblock A random riemannian metric for probabilistic shortest-path tractography.
\newblock In Nassir Navab, Joachim Hornegger, William~M. Wells, and Alejandro Frangi, editors, \emph{Medical Image Computing and Computer-Assisted Intervention -- MICCAI 2015}, pages 597--604, Cham, 2015. Springer International Publishing.
\newblock ISBN 978-3-319-24553-9.

\bibitem[Hsu(2002)]{hsustochastic}
E.P. Hsu.
\newblock \emph{Stochastic Analysis on Manifolds}.
\newblock Contemporary Mathematics. American Mathematical Soc., 2002.
\newblock ISBN 9780821883884.
\newblock URL \url{https://books.google.dk/books?id=2NM0Z7svRmEC}.

\bibitem[Jensen and Sommer(2022)]{jensensommer2022}
Mathias~Højgaard Jensen and Stefan Sommer.
\newblock Mean estimation on the diagonal of product manifolds.
\newblock \emph{Algorithms}, 15\penalty0 (3), 2022.
\newblock ISSN 1999-4893.
\newblock \doi{10.3390/a15030092}.
\newblock URL \url{https://www.mdpi.com/1999-4893/15/3/92}.

\bibitem[Jensen and Sommer(2023)]{jensen2023simulation}
Mathias~Højgaard Jensen and Stefan Sommer.
\newblock Simulation of conditioned semimartingales on riemannian manifolds, 2023.

\bibitem[Jørgensen(1975)]{Jrgensen1975TheCL}
Erik Jørgensen.
\newblock The central limit problem for geodesic random walks.
\newblock \emph{Zeitschrift f{\"u}r Wahrscheinlichkeitstheorie und Verwandte Gebiete}, 32:\penalty0 1--64, 1975.

\bibitem[Karcher(1977)]{Karcher1977RiemannianCO}
Hermann Karcher.
\newblock Riemannian center of mass and mollifier smoothing.
\newblock \emph{Communications on Pure and Applied Mathematics}, 30:\penalty0 509--541, 1977.

\bibitem[Lou et~al.(2021)Lou, Katsman, Jiang, Belongie, Lim, and Sa]{lou2021differentiating}
Aaron Lou, Isay Katsman, Qingxuan Jiang, Serge Belongie, Ser-Nam Lim, and Christopher~De Sa.
\newblock Differentiating through the fr\'echet mean, 2021.

\bibitem[Meng et~al.(2021)Meng, Song, Li, and Ermon]{meng2021estimating}
Chenlin Meng, Yang Song, Wenzhe Li, and Stefano Ermon.
\newblock Estimating high order gradients of the data distribution by denoising, 2021.

\bibitem[Papaspiliopoulos and Roberts(2012)]{Papaspiliopoulos2012}
Omiros Papaspiliopoulos and G.~Roberts.
\newblock \emph{Importance sampling techniques for estimation of diffusion models}, pages 311--340.
\newblock 01 2012.

\bibitem[Pennec(2006)]{pennec2006statriemann}
Xavier Pennec.
\newblock Intrinsic statistics on riemannian manifolds: Basic tools for geometric measurements.
\newblock \emph{Journal of Mathematical Imaging and Vision}, 25:\penalty0 127--154, 07 2006.
\newblock \doi{10.1007/s10851-006-6228-4}.

\bibitem[Salimans and Ho(2021)]{learn_log_p}
Tim Salimans and Jonathan Ho.
\newblock Should {EBM}s model the energy or the score?
\newblock In \emph{Energy Based Models Workshop - ICLR 2021}, 2021.
\newblock URL \url{https://openreview.net/forum?id=9AS-TF2jRNb}.

\bibitem[Sommer(2018)]{sommer2018infinitesimal}
Stefan Sommer.
\newblock An infinitesimal probabilistic model for principal component analysis of manifold valued data, 2018.

\bibitem[Sommer et~al.(2017)Sommer, Arnaudon, Kuhnel, and Joshi]{sommer2017bridge}
Stefan Sommer, Alexis Arnaudon, Line Kuhnel, and Sarang Joshi.
\newblock Bridge simulation and metric estimation on landmark manifolds, 2017.

\bibitem[Song et~al.(2019)Song, Garg, Shi, and Ermon]{song2019sliced}
Yang Song, Sahaj Garg, Jiaxin Shi, and Stefano Ermon.
\newblock Sliced score matching: A scalable approach to density and score estimation, 2019.

\bibitem[Song et~al.(2021)Song, Durkan, Murray, and Ermon]{song2021maximum}
Yang Song, Conor Durkan, Iain Murray, and Stefano Ermon.
\newblock Maximum likelihood training of score-based diffusion models, 2021.

\bibitem[Sotiropoulos et~al.(2013)Sotiropoulos, Moeller, Jbabdi, Xu, Andersson, Auerbach, Yacoub, Feinberg, Setsompop, Wald, Behrens, Ugurbil, and Lenglet]{dti_ref2}
S.~N. Sotiropoulos, S.~Moeller, S.~Jbabdi, J.~Xu, J.~L. Andersson, E.~J. Auerbach, E.~Yacoub, D.~Feinberg, K.~Setsompop, L.~L. Wald, T.~E.~J. Behrens, K.~Ugurbil, and C.~Lenglet.
\newblock Effects of image reconstruction on fiber orientation mapping from multichannel diffusion mri: Reducing the noise floor using sense.
\newblock \emph{Magnetic Resonance in Medicine}, 70\penalty0 (6):\penalty0 1682--1689, 2013.
\newblock \doi{https://doi.org/10.1002/mrm.24623}.
\newblock URL \url{https://onlinelibrary.wiley.com/doi/abs/10.1002/mrm.24623}.

\bibitem[Tosi et~al.(2014)Tosi, Hauberg, Vellido, and Lawrence]{tosi2014metrics}
Alessandra Tosi, Søren Hauberg, Alfredo Vellido, and Neil~D. Lawrence.
\newblock Metrics for probabilistic geometries, 2014.

\bibitem[{Van Essen} et~al.(2013){Van Essen}, Smith, Barch, Behrens, Yacoub, and Ugurbil]{dti_ref3}
David~C. {Van Essen}, Stephen~M. Smith, Deanna~M. Barch, Timothy~E.J. Behrens, Essa Yacoub, and Kamil Ugurbil.
\newblock The wu-minn human connectome project: An overview.
\newblock \emph{NeuroImage}, 80:\penalty0 62--79, 2013.
\newblock ISSN 1053-8119.
\newblock \doi{https://doi.org/10.1016/j.neuroimage.2013.05.041}.
\newblock URL \url{https://www.sciencedirect.com/science/article/pii/S1053811913005351}.
\newblock Mapping the Connectome.

\bibitem[Varadhan(1967)]{varadhan_heat}
S.~R.~S. Varadhan.
\newblock On the behavior of the fundamental solution of the heat equation with variable coefficients.
\newblock \emph{Communications on Pure and Applied Mathematics}, 20\penalty0 (2):\penalty0 431--455, 1967.
\newblock \doi{https://doi.org/10.1002/cpa.3160200210}.
\newblock URL \url{https://onlinelibrary.wiley.com/doi/abs/10.1002/cpa.3160200210}.

\bibitem[Vincent(2011)]{pascal_dsm}
Pascal Vincent.
\newblock A connection between score matching and denoising autoencoders.
\newblock \emph{Neural Computation}, 23\penalty0 (7):\penalty0 1661--1674, 2011.
\newblock \doi{10.1162/NECO_a_00142}.

\bibitem[Wang et~al.(2020)Wang, Cheng, Li, Zhu, and Zhang]{wang2020wasserstein}
Ziyu Wang, Shuyu Cheng, Yueru Li, Jun Zhu, and Bo~Zhang.
\newblock A wasserstein minimum velocity approach to learning unnormalized models, 2020.

\bibitem[Zhao and Song(2018)]{zhaohksm}
Chenchao Zhao and Jun~S. Song.
\newblock Exact heat kernel on a hypersphere and its applications in kernel svm.
\newblock \emph{Frontiers in Applied Mathematics and Statistics}, 4, 2018.
\newblock ISSN 2297-4687.
\newblock \doi{10.3389/fams.2018.00001}.
\newblock URL \url{https://www.frontiersin.org/articles/10.3389/fams.2018.00001}.

\bibitem[Zhu et~al.(2019)Zhu, Liu, Lei, and Li]{aflw2000}
Xiangyu Zhu, Xiaoming Liu, Zhen Lei, and Stan~Z. Li.
\newblock Face alignment in full pose range: A 3d total solution.
\newblock \emph{IEEE Transactions on Pattern Analysis and Machine Intelligence}, 41\penalty0 (1):\penalty0 78–92, January 2019.
\newblock ISSN 1939-3539.
\newblock \doi{10.1109/tpami.2017.2778152}.
\newblock URL \url{http://dx.doi.org/10.1109/TPAMI.2017.2778152}.

\bibitem[Ziezold(1977)]{Ziezold1977}
Herbert Ziezold.
\newblock \emph{On Expected Figures and a Strong Law of Large Numbers for Random Elements in Quasi-Metric Spaces}, pages 591--602.
\newblock Springer Netherlands, Dordrecht, 1977.
\newblock ISBN 978-94-010-9910-3.
\newblock \doi{10.1007/978-94-010-9910-3_63}.
\newblock URL \url{https://doi.org/10.1007/978-94-010-9910-3_63}.

\end{thebibliography}
